\documentclass[twocolumn,showpacs,aps,superscriptaddress]{revtex4-1}
\include{def}
\usepackage{amsmath}
\usepackage{lineno}
\usepackage{graphicx}
\usepackage{xcolor}
\usepackage{xcolor}
\colorlet{RED}{red}
\usepackage{longtable}
\usepackage{multirow}
\usepackage{makecell}

\def\JPG{J.Phys.G}
\def\JHEP{J. High Energ. Phys.}
\def\EPJ{Eur.Phys.J.}
\def\ARNPS{Ann.Rev.Nucl.Part.Sci.}
\def\NPA{Nucl.Phys.A}
\def\NPB{Nucl.Phys.B}

\def\PLB{Phys.Lett.B}

\def\PRL{Phys.Rev.Lett.}
\def\PRC{Phys.Rev.C}
\def\PRD{Phys.Rev.D}

\def\EPJ{Eur.Phys.J.}

\def\APP{Acta Phys. Pol.}
\def\CPC{Comput. Phys. Commun.}
\def\be{\begin{equation}}
\def\ee{\end{equation}}

\begin{document}

\setlength\LTcapwidth{\linewidth}
\title{Considerations on the suppression of charged particles and $\pi^0$ in\\
 high energy heavy ion collisions}

\author{M.Petrovici}
\affiliation{National Institute for Physics and Nuclear Engineering - IFIN-HH\\
Hadron Physics Department\\
Bucharest - Romania}
\affiliation{Faculty of Physics, University of Bucharest}
\author{A.Lindner}
\affiliation{National Institute for Physics and Nuclear Engineering - IFIN-HH\\
Hadron Physics Department\\
Bucharest - Romania}
\affiliation{Faculty of Physics, University of Bucharest} 
\author{A.Pop} 
\affiliation{National Institute for Physics and Nuclear Engineering - IFIN-HH\\
Hadron Physics Department\\
Bucharest - Romania}
\date{\today}

\begin{abstract}
Experimental results related to charged particle and $\pi^{0}$ suppression obtained at the
 Relativistic Heavy Ion Collider (RHIC) at Brookhaven for Au-Au (Cu-Cu) collisions and at the Large Hadron Collider (LHC) at CERN for Pb-Pb (Xe-Xe) collisions are 
 compiled in terms of the usual nuclear modification factors, $R_{AA} $ and $R_{CP}$, and of the newly introduced $R^{N}_{AA}$ and $R^{N}_{CP}$ as a function of $\langle N_{part} \rangle$  and $\langle dN_{ch}/d\eta\rangle$. The $R^{N}_{AA}$ and $R^{N}_{CP}$ are calculated as the ratios of the $p_{T}$ spectra in each centrality bin, to the spectrum in proton-proton minimum bias collisions, or to the spectrum in a peripheral bin, respectively, each of them normalised to the corresponding charged particle density.
  The studies are focused on a $p_T$ range in the region of maximum suppression evidenced in the experiments. The $R_{AA}$ scaling as a function of $\langle N_{part} \rangle$ and $\langle dN_{ch}/d\eta \rangle$ is discussed.  The core contribution to $R_{AA}$ is presented. The difference in  $R_{AA}$ relative to the difference in particle density per unit of rapidity and unit of overlapping area ($\langle dN/dy \rangle /S_{\perp}$) and the Bjorken energy density times the interaction time ($\varepsilon_{Bj}\cdot\tau$) between top RHIC and LHC energies indicate a suppression saturation at LHC energies. Considerations on the missing suppression in high charged particle multiplicity events for pp collisions at 7 TeV are presented. $R^N_{CP}$ for the same systems and energies shows  a linear scaling as a function of $\langle N_{part}\rangle$. 
While (1-$R_{AA}$)/$\langle dN/dy \rangle$ shows an exponential decrease with 
$(\langle dN/dy \rangle/S_{\perp})^{1/3}$, (1-$R_{AA}^N$)/$\langle dN/dy \rangle$ is independent on $(\langle dN/dy \rangle/S_{\perp})^{1/3}$ for $(\langle dN/dy \rangle/S_{\perp})^{1/3}\geq$2.1 particles$/fm^{2/3}$. 
The trends of $R_{CP}$ and $R^N_{CP}$ for charged particles as a function of $\sqrt{s_{NN}}$, measured at RHIC in Au-Au collisions and at LHC in Pb-Pb collisions, show a suppression that becomes larger from 
$\sqrt{s_{NN}}$ = 39 GeV up to $\sqrt{s_{NN}}$=200 GeV, followed by a saturation up to the highest energy of $\sqrt{s_{NN}}$ =5.02 TeV in Pb-Pb collisions. The $\sqrt{s_{NN}}$ dependences of $R_{AA}^{\pi^0}$ and $(R_{AA}^N)^{\pi^0}$ in the same $p_T$ ranges and for the very central collisions show the same trend. 
A clear change in the dependence of $(1-R_{AA}^{\pi^0})/\langle dN/dy \rangle$ for the most central collisions as a function of collision energy is evidenced in the region of 
$\sqrt{s_{NN}}$ =62.4 - 130 GeV.
\end{abstract}

\maketitle

\section{Introduction}
 Detailed studies of different observables in heavy ion collisions at RHIC \cite{BRAHMS05, PHOBOS05, STAR05, PHENIX05, PHENIXB, STARB} support theoretical predictions pioneered more than 40 years ago \cite{Coll, Cha, Shu, Chi} that at large densities and temperatures of the fireballs produced at these energies, the matter is deconfined into its basic constituents, quarks and gluons. Obviously, such studies are rather difficult given that the produced fireballs are highly non-homogeneous, have a small size and are highly unstable, since their dynamical evolution plays an important role. One of the powerful tools used to diagnose the properties of such a deconfined object is the study of the energy loss of partons traversing the deconfined matter \cite{Bjorken82}. 
 Within QCD based models, the energy loss of a parton traversing deconfined matter is due to collisional or radiative processes. Collisional energy loss due to elastic parton collisions is expected to scale linearly with the path length \cite{dentbetz}. Radiative energy loss occurs via inelastic processes where a hard parton radiates a gluon. Soft interactions of partons with the deconfined medium can also induce gluon radiation \cite{dEnterria10}. Radiative energy loss is expected to grow quadratically with the path length \cite{Renk07}. There are quite a few theoretical approaches to describe the parton energy loss in expanding deconfined matter \cite{Baier97, Gyulassy00, Baier01, Arleo02, Muller05, Djordjevic08, Casalderrey-Solana15, Burke14, Betz14, Arleo18}. 
 However, a proper description of the parton energy loss in the non-equilibrium expanding deconfined matter for the intermediate $p_T$ range remains a challenging task. The predicted suppression at LHC energies turned out to be overestimated, once the experimental information became available. A comprehensive analysis within the CUJET/CIBJET framework recently published \cite{Shi19}, indicates, similar to the results of the JET Collaboration \cite{Burke14}, a maximum in $\hat{q}/T^3$ as a function of temperature around the critical temperature ($T_c$), followed by a decrease towards temperatures reached at LHC energies. 
Some considerations on the charged particle  and $\pi^0$ suppression at RHIC and LHC energies are presented in this paper.
Section II is a short presentation of the quantities estimated in the Glauber Monte Carlo (MC) model used in the next sections.
 A review of the charged particle suppression dependence on 
$\langle N_{part} \rangle$ and $\langle dN_{ch}/d\eta \rangle$, the core-corona effect and the dependence on particle density per unit of rapidity and unit of overlapping area ($\langle dN/dy \rangle/S_{\perp}$), a measure of the entropy density and thus of temperature \cite{Vogt} and the Bjorken energy density times the interaction time ($\varepsilon_{Bj}\cdot\tau$) for Cu-Cu and Au-Au at the top RHIC energy and for Xe-Xe and Pb-Pb at LHC energies, are presented in Section III. Section IV is dedicated to similar studies, using $\langle dN_{ch}/d\eta \rangle^{AA}/\langle dN_{ch}/d\eta\rangle^{pp}$ instead of $\langle N_{bin} \rangle$ in a model independent estimation of suppression, namely $R_{AA}^N$, defined later in this section \cite{Petrovici17_aip}. In Section V, similar considerations for the corresponding relative suppression, $R_{CP}$ and $R_{CP}^N$ are presented.
(1-$R_{AA}$)/$\langle dN/dy \rangle$ and (1-$R_{AA}^N$)/$\langle dN/dy \rangle$ dependences as a function of
$(\langle dN/dy \rangle/S_{\perp})^{1/3}$ are presented in Section VI. 
 The collision energy dependence of $R_{CP}$, $R_{CP}^N$ for charged particles and $R_{AA}$, $R_{AA}^N$ for $\pi^0$ is discussed in Section VII. Conclusions are presented in Section VIII.
 
\section{Glauber Monte Carlo estimates}
\begin{figure}[t!]
\centering
\includegraphics[scale=0.48]{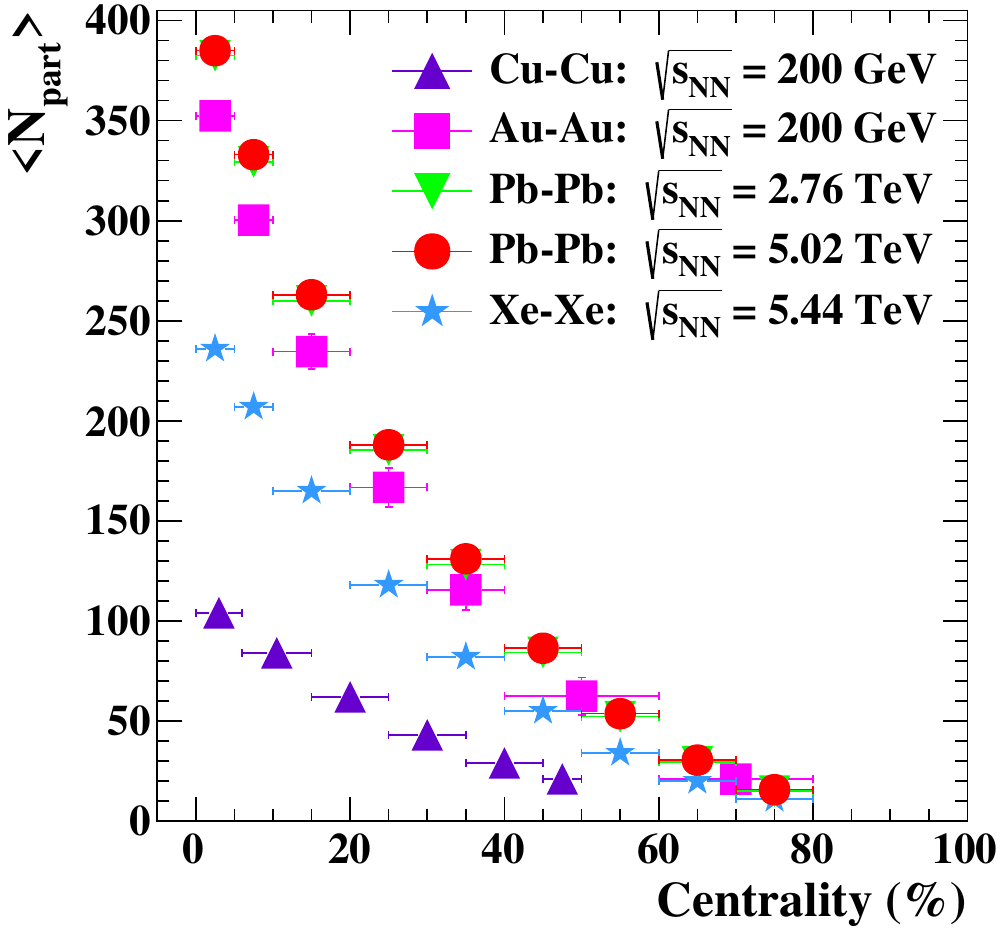}
\caption{The average number of participating nucleons $\langle N_{part} \rangle$ as a function of centrality for Cu-Cu, Au-Au collisions at $\sqrt{s_{NN}}$ = 200 GeV, for Xe-Xe at \protect\linebreak $\sqrt{s_{NN}}$ = 5.44 TeV and for Pb-Pb at $\sqrt{s_{NN}}$ = 2.76 and 5.02 TeV.}
\label{fig-1}
\end{figure}
\begin{figure}[h!]
\centering
\includegraphics[scale=0.48]{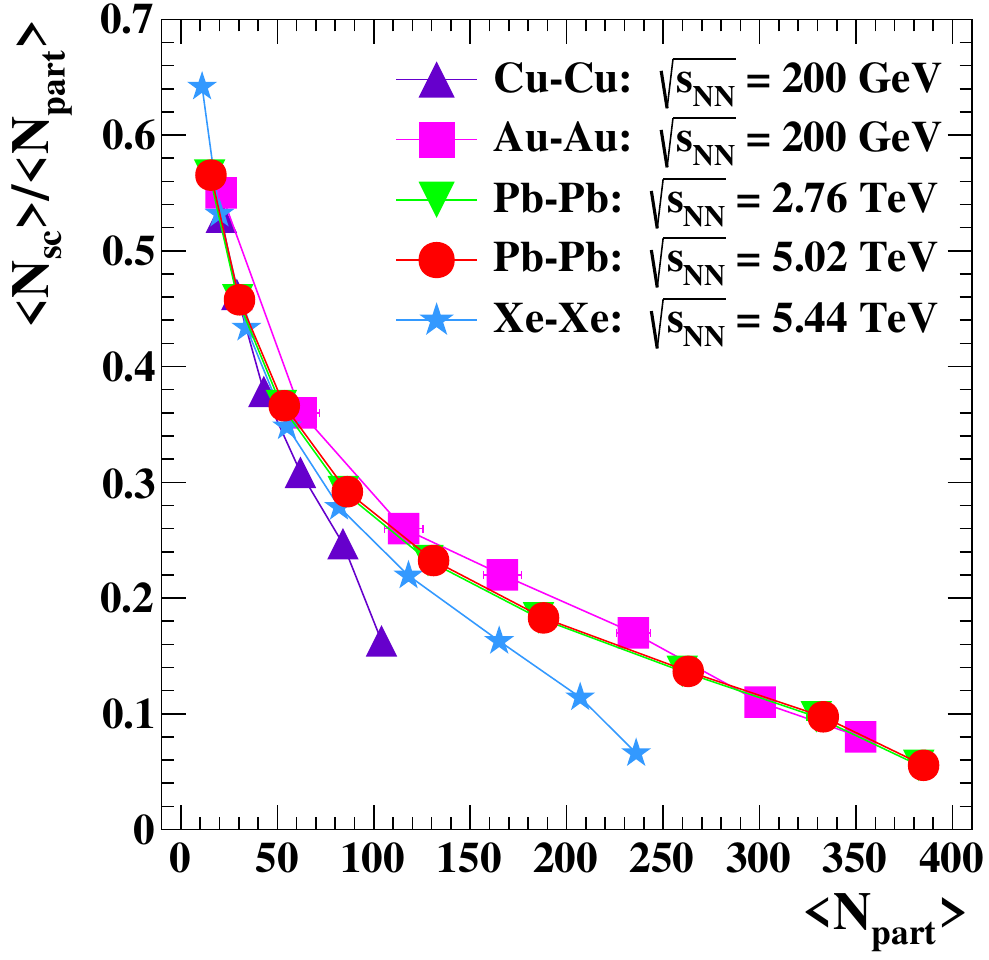}
\caption{Ratio of the average number of nucleons undergoing single collisions to the average number of participating nucleons ($\langle N_{sc} \rangle/\langle N_{part} \rangle$), as a function of the average number of participating nucleons ($\langle N_{part} \rangle$) estimated within the Glauber MC model.}
\label{fig-2}
\end{figure}

   The Glauber MC model \cite{Glauber55, Franco66, Miller07, Rybcz2014} was used to estimate in a unitary manner various quantities characteristic to the initial state in A-A collisions: number of participants, number of collisions, number of nucleons undergoing a single collision and the transverse overlapping areas in centrality bins. In a core-corona picture \cite{Rybcz2014}, the core quantities were estimated for wounded nucleons suffering more than a single collision. The calculations have been done in the hard sphere 
wounding prescription \cite{Rybcz2014}. For the nuclear density profile of the colliding nuclei, a Woods-Saxon
distribution was considered:
\begin{equation} 
\rho(r)=\frac{1}{1+exp(\frac{r-r_{0}}{a})}
\end{equation}
with a = 0.535 fm, $r_{0}$ = 6.5 fm for the Au nucleus \cite{STAR_dNchdeta},
a = 0.546 fm, $r_{0}$ = 6.62 fm for the Pb nucleus \cite{Abelev044909}
a = 0.57 fm, $r_{0}$ = 5.42 fm for the Xe nucleus \cite{Loiz} and a = 0.596 fm, $r_{0}$ = 4.2 fm for the Cu nucleus \cite{Loiz}.
Within
the hard sphere approach, the nucleons are considered to collide
if the relative transverse distance $d\leq\sqrt{\frac{\sigma_{pp}}{\pi}}$. The nucleon - nucleon inelastic cross section, $\sigma_{pp}$, at a given collision energy, was taken as specified in Refs. \cite{STAR_dNchdeta, Abelev044909, XeXecent, PbPb_502_dNchdeta}.
The centrality dependence of the overlapping area, $S_{\perp}^{var}$ is considered to be proportional to the quantity
$S \propto\sqrt{\langle \sigma_{x}^{2} \rangle\langle \sigma_{y}^{2} \rangle-\langle \sigma_{xy}^{2} \rangle}$. $\sigma_{x}^{2}$, $\sigma_{y}^{2}$ are the variances, and
$\sigma_{xy}$ is the co-variance of the participant distributions in the
transverse plane, per event \cite{Alver08}. They were averaged 
over many events.
The centrality dependent values were rescaled in such a way as to equalize the geometrical area (calculated as in \cite{Petrovici18, PetAIP}) and 
$S_{\perp}^{var}$ in the case of the complete overlap of the nuclei (b = 0 fm).
After generating a large number of events for the minimum bias (MB) collisions, they were sorted in centrality classes according to the impact parameter distribution. Calculations of the quantities of interest have been done in each centrality class. 
The results of the calculations for various systems and energies were presented in Refs. \cite{Petrovici18, PetAIP} and Table I of this paper. 
The obtained number of participants and number of collisions are in good agreement, within the error bars, with the same quantities listed in different experimental publications. 
In Figure \ref{fig-1}, the average number of participating nucleons ($\langle N_{part} \rangle$)  \cite{Alver06, STAR_AuAu_200, XeXe_data, PbPb_502_dNchdeta, Petrovici18}
as a function of centrality obtained within the Glauber
MC approach is shown. As
can be seen, the difference in $\langle N_{part} \rangle$ at a given centrality, for colliding systems with different sizes and incident
energies, is increasing from peripheral towards central
collisions. Figure \ref{fig-2} shows the average number of
nucleons undergoing single collisions relative to the average number of participating nucleons ($\langle N_{sc} \rangle/\langle N_{part} \rangle$).
As expected, $\langle N_{sc} \rangle/\langle N_{part} \rangle$ has large values at low
$\langle N_{part} \rangle$, the system size and collision energy dependence
being rather small. With increasing $\langle N_{part} \rangle$ towards
very central collisions, although the percentage of nucleons undergoing single collisions decreases, the difference between the various systems becomes significant.

\begin{longtable}{|c|c|c|c|c|c|}
\caption{The percentage of nucleons that suffer more than a single collision ($f_{core}$), the overlapping surface of the colliding nuclei ($S_{\perp}^{var}$) and the overlapping surface corresponding to the core contribution (($S_{\perp}^{var})^{core}$) for Cu-Cu and Xe-Xe colliding systems at the corresponding collision energies and centralities.}\\
\hline
\bf {System} &  $\bf \sqrt{s_{NN}}$ & \bf Cen. & \bf $f_{core}$ & $\bf S_{\perp}^{var}$ ) &  $\bf (S_{\perp}^{var})^{core}$\\ 
     & (GeV) & (\%)  &   &  ($fm^{2}$) & ($fm^{2}$)\\
\hline
\hline
	   \multirow{4}{*}{Cu-Cu} & \multirow{4}{*}{200} 
          &  0-10 & 0.81$\pm$0.00 & 67.9$\pm$0.5 & 51.8$\pm$0.4\\
       &  & 10-30 & 0.69$\pm$0.00 & 53.4$\pm$0.4 & 36.1$\pm$0.3\\
	   &  & 30-50 & 0.55$\pm$0.00 & 38.3$\pm$0.3 & 23.3$\pm$0.2\\
	   &  & 50-70 & 0.38$\pm$0.01 & 24.7$\pm$0.2 & 13.2$\pm$0.1\\
	   \hline
	   \hline
\multirow{9}{*}{Xe-Xe} & \multirow{9}{*}{5440} 
          &  0-5  & 0.93$\pm$0.00 & 124.1$\pm$0.6 & 105.3$\pm$0.5\\
       &  &  5-10 & 0.89$\pm$0.00 & 114.9$\pm$0.6 & 91.3$\pm$0.5\\
	   &  & 10-20 & 0.84$\pm$0.00 & 100.6$\pm$0.5 & 74.9$\pm$0.4\\
	   &  & 20-30 & 0.78$\pm$0.00 & 83.7$\pm$0.5 & 57.9$\pm$0.3\\
	   &  & 30-40 & 0.72$\pm$0.00 & 69.3$\pm$0.4 & 44.7$\pm$0.2\\
	   &  & 40-50 & 0.65$\pm$0.00 & 57.1$\pm$0.3 & 34.2$\pm$0.2\\
	   &  & 50-60 & 0.57$\pm$0.00 & 45.9$\pm$0.3 & 25.5$\pm$0.1\\
	   &  & 60-70 & 0.47$\pm$0.01 & 35.4$\pm$0.2 & 18.2$\pm$0.1\\
	   &  & 70-80 & 0.36$\pm$0.01 & 24.8$\pm$0.2 & 10.9$\pm$0.1\\
	   \hline
\end{longtable} 

\begin{figure}[h!]
\centering
\includegraphics[scale=0.48]{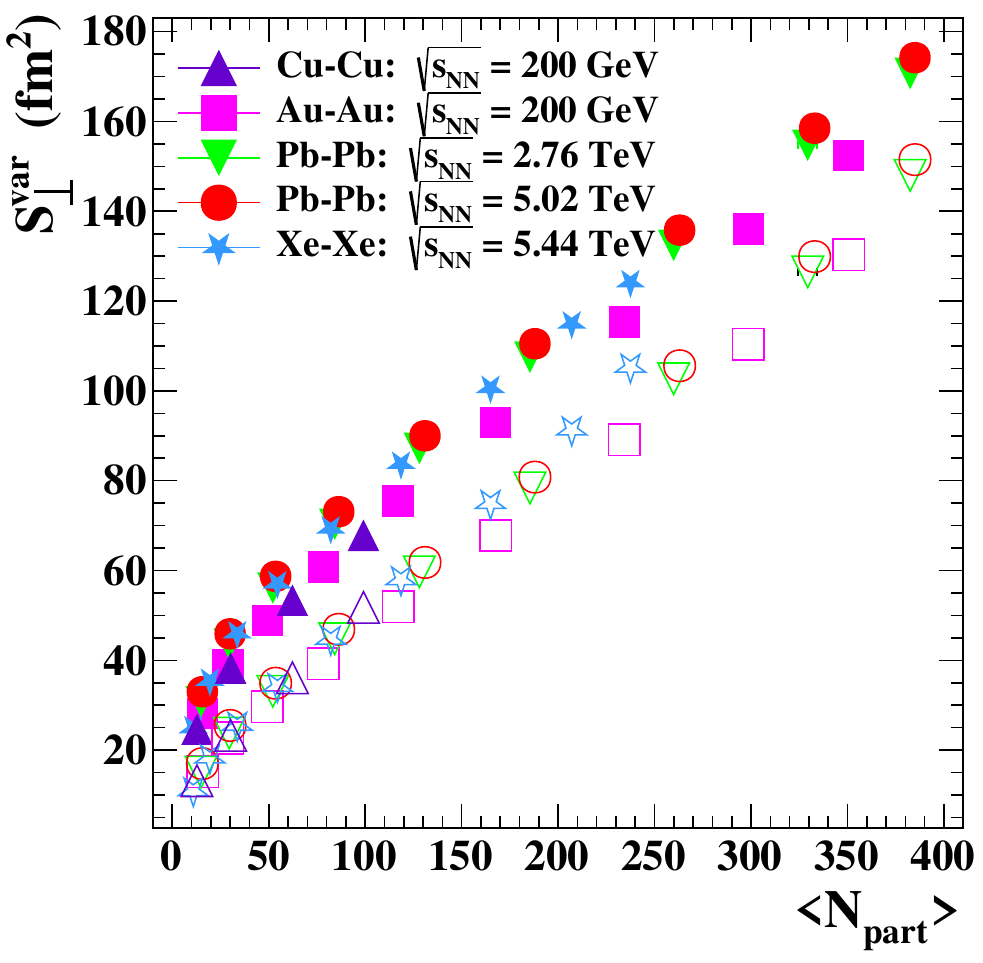}
\caption{ The overlapping area ($S_{\perp}^{var}$) as a function of $\langle N_{part} \rangle$ 
corresponding to total (full symbols) and core (open symbols) wounded nucleons.}
\label{fig-3}
\end{figure}

Figure \ref{fig-3} shows the overlapping area, $S_{\perp}^{var}$, as a function of $\langle N_{part} \rangle$ for the total and core contribution. In this paper we decided to use 
$S_{\perp}^{var}$, similar to what was used to estimate the Bjorken energy density at LHC energies \cite{Adam16}. It will be simply written $S_{\perp}$ from now on.

\section{$R_{AA}$ ($5<p_T<8$ $GeV/c$) : $\langle N_{part} \rangle$ dependence}
 Usually, the comparisons among different systems and different collision energies in terms of the nuclear modification factor, $R_{AA}$, are done as a function of collision centrality. $R_{AA}$ is defined as:   
 
\begin{equation}
R_{AA}=\frac{(\frac{d^{2}N}{d\eta dp_{T}})^{cen}}{\langle N_{bin}\rangle\cdot (\frac{d^{2}N}{d\eta dp_{T}})^{pp,MB}} 
\end{equation}  
\noindent
where the transverse momentum distribution of a certain particle measured in A-A collisions for a given centrality (cen) is divided by the pp MB $p_{T}$ distribution 
of that particle at the same energy, multiplied by the number of binary collisions calculated based on the Glauber MC model.
\begin{figure*}[t!]
\centering
\includegraphics[width=0.95\linewidth]{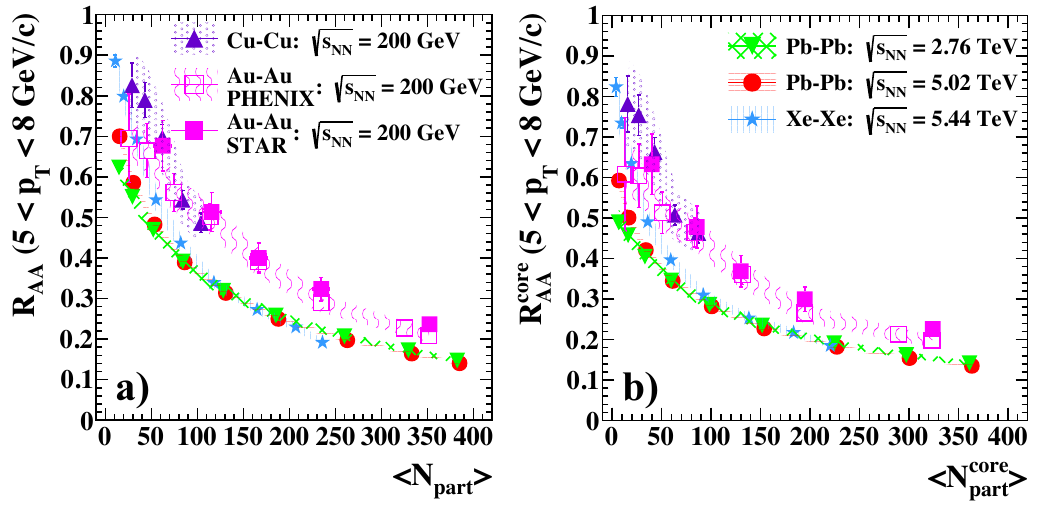}
\caption{$R_{AA}$ in the $5<p_T<8$ GeV/c region as a function of the average number of nucleons $\langle N_{part} \rangle$ for charged particles; a) experimental values, b) core contribution, see the text.}
\label{fig-4}
\end{figure*}
Because of the $\langle N_{part} \rangle$ dependence on centrality in Figure \ref{fig-1}, a study of the suppression phenomena in relativistic heavy ion collisions as a function of system size and collision energy is better done in terms of $\langle N_{part} \rangle$, instead of centrality. 
At $\sqrt{s_{NN}}$=200 GeV, the same values of charged particle $R_{AA}$ as a function of $\langle N_{part} \rangle$ for different bins in $p_T$, for two very different colliding symmetric systems Au-Au \cite{Adler04} and Cu-Cu \cite{Alver06}, were evidenced. A similar scaling was also observed for a lower collision energy, i.e. $\sqrt{s_{NN}}$=62.4 GeV \cite{Back05}. 
 Such a dependence was studied for pions and protons, for 5$<p_T<$8 GeV/c and 5$<p_T<$6 GeV/c respectively, in Cu-Cu and Au-Au collisions
at $\sqrt{s_{NN}}$=200 GeV, by the STAR Collaboration \cite{Abelev10}, where a good scaling of $R_{AA}^{\pi^{+}+\pi^{-}}$ as a function of $\langle N_{part} \rangle$ for the two systems was seen. The PHENIX Collaboration has shown that in Au-Au collisions at $\sqrt{s_{NN}}$=62.4 GeV and 200 GeV, the $R_{AA}$ of $\pi^0$ for $p_T>$ 6 GeV/c has the same value as a function of $\langle N_{part} \rangle$ \cite{Adare12}. 
 At the LHC energies, the CMS Collaboration presented a similar scaling for Xe-Xe at $\sqrt{s_{NN}}$=5.44 TeV and Pb-Pb at $\sqrt{s_{NN}}$=5.02 TeV \cite{Sirunyan18} with the remark that the $R_{AA}$ for Xe-Xe was obtained using the $p_T$ spectrum from MB pp collisions at $\sqrt{s}$=5.02 TeV. 
 Aside from Cu-Cu at $\sqrt{s_{NN}}$=200 GeV, where the $p_T$ spectra were obtained for the 0.2 $< \eta <$ 1.4 pseudorapidity range, all the other results were obtained for a symmetric cut relative to $\eta$=0. 
 
 Suppression studies at LHC energies up to very large $p_T$ values \cite{Chatrchyan12, Abelev13, Aad15}, for charged particles, evidence a maximum suppression in the 5-8 GeV/c $p_T$ range, for a given centrality. While the absolute value of the maximum suppression depends on centrality, its position is
 in the same region of $p_T$.  Although at RHIC energies the measured $p_T$ range is much smaller than the region where the $R_{AA}$ starts to increase, based on the larger range in $p_T$ for $\pi^0$ \cite{Adare12}, one could conclude that the maximum suppression for different centralities is in the same range of $p_T$, i.e. 5-8 GeV/c. This is the main reason to focus the present considerations on the suppression phenomena in this $p_T$ range.  

 Using the latest results obtained at RHIC for Cu-Cu and Au-Au collisions at $\sqrt{s_{NN}}$=200 GeV \cite{Alver06, STAR_AuAu_200, Adler04}, at LHC for Xe-Xe at $\sqrt{s_{NN}}$=5.44 TeV \cite{XeXe_data} and  Pb-Pb at $\sqrt{s_{NN}}$=2.76 and 5.02 TeV \cite{PbPb_Raa}, we obtained the mean values of $R_{AA}$, averaged over the $5<p_T<8$ GeV/c region, presented in Figure \ref{fig-4}a. $R_{AA}$ scales as a function of $\langle N_{part} \rangle$ at RHIC ($\sqrt{s_{NN}}$=200 GeV) and LHC energies, separately, as it was shown in the above mentioned papers. Within the error bars, a small difference, i.e. a slightly larger suppression is observed for central Cu-Cu and Xe-Xe collisions relative to Au-Au and Pb-Pb respectively, at the same $\langle N_{part} \rangle$. The highlighted areas represent the systematic uncertainties, while the error bars are the statistical uncertainties, for the cases where they have been reported separately (Pb-Pb at $\sqrt{s_{NN}}$=2.76 and 5.02 TeV, Xe-Xe at $\sqrt{s_{NN}}$=5.44 TeV, Au-Au (PHENIX) and  Cu-Cu at $\sqrt{s_{NN}}$=200 GeV), while in the case of Au-Au (STAR) the error bars represent the square root of statistical and systematic uncertainties added in quadrature. 
Another aspect worth being considered is the so called core-corona effect \cite{Becattini04, Bozek05, Werner07, Steinheimer11, Becattini08, Becattini09, Aichelin09, Aichelin10_2, Bozek09, Schreiber12, Gemard14, Petrovici17} on the suppression estimate. 
The contribution to the $p_T$ spectra in A-A collisions from a nucleon suffering a single collision is similar with the spectra from  pp MB collisions at the same energy. Therefore, one should first correct the experimental spectra of A-A collisions with the contribution coming from single binary collisions (corona) in order to obtain the spectra of the core:

\begin{equation}
\begin{split}
&(\frac{d^{2}N}{d\eta dp_{T}})^{cen,core}=(\frac{d^{2}N}{d\eta dp_{T}})^{cen,measured} - \\
&-(\frac{d^{2}N}{d\eta dp_{T}})^{pp,MB,measured}\cdot\frac{\langle N_{part} \rangle^{cen}}{2}\cdot(1-f^{cen}_{core}),
\end{split}
\end{equation}

\noindent
where $f^{cen}_{core}=\langle N_{part}^{core} \rangle^{cen}/\langle N_{part} \rangle^{cen}$.
 The suppression due to the core of the fireball, $R^{core}_{AA}$:
\begin{equation}
R_{AA}^{core}=\frac{(\frac{d^{2}N}{d\eta dp_{T}})^{cen,core}}{\langle N_{bin}^{core}\rangle^{cen}\cdot (\frac{d^{2}N}{d\eta dp_{T}})^{pp,MB}},
\end{equation}   
where
\begin{equation}
\langle N_{bin}^{core}\rangle^{cen}= \langle N_{bin} \rangle^{cen}-\frac{\langle N_{part}\rangle^{cen}}{2}\cdot(1-f_{core}^{cen})
\end{equation}
is presented in Figure \ref{fig-4}b as a function of $\langle N_{part}^{core} \rangle$. In the simple image of a net core-corona separation, the figure shows the core contribution extracted from the experimental data for the different centrality classes. 
\begin{figure}
\centering
\includegraphics[scale=0.5]{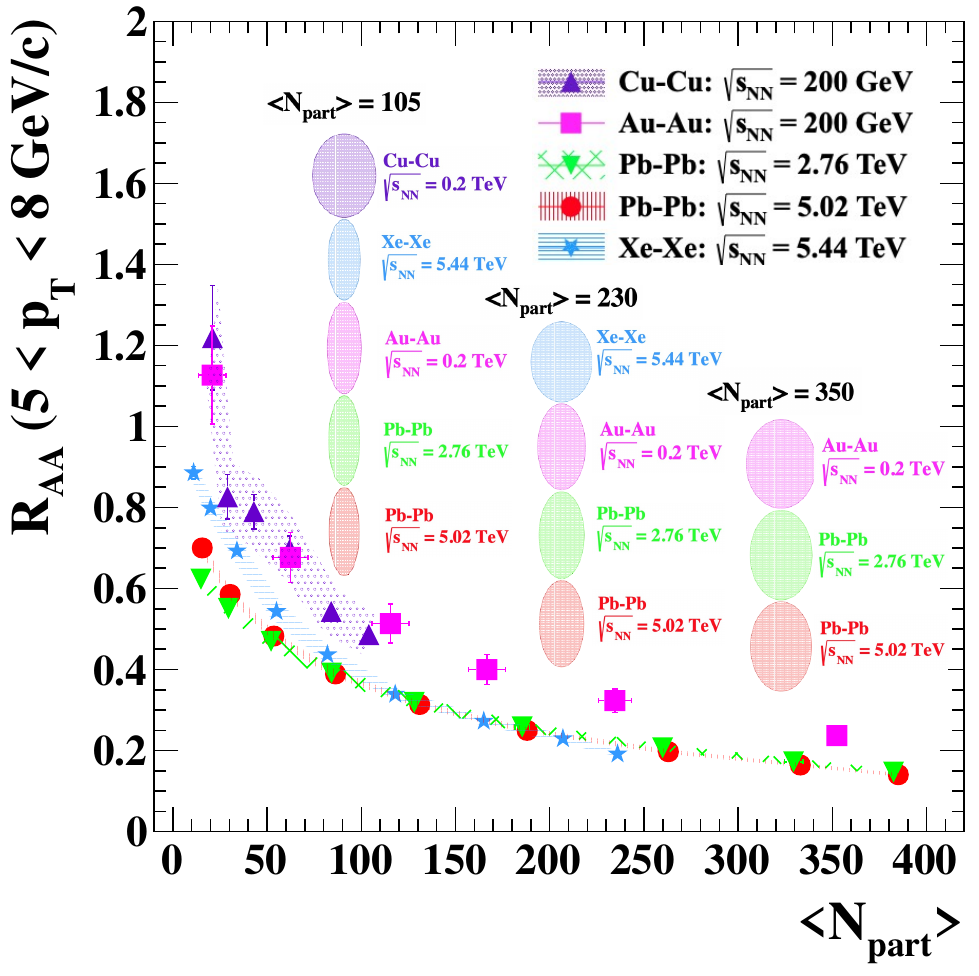}
\caption{The same as Figure \ref{fig-4}a with the shape of the overlapping area $S_{\perp}$ at different values of $\langle N_{part} \rangle$ estimated within the Glauber MC approach.}
\label{fig-5}
\end{figure}

The suppression is increased in peripheral collisions by $\approx$10-20\% and the values for the most central Cu-Cu and Xe-Xe collisions are the same as for Au-Au and Pb-Pb collisions, respectively, for the same $\langle N_{part}^{core} \rangle$. The suppression for Cu-Cu and Au-Au is the same at the same collision energy $(\sqrt{s_{NN}}$=200 GeV). At the LHC energies, the suppression in Pb-Pb collisions at $\sqrt{s_{NN}}$=2.76 TeV is the same as at 
$\sqrt{s_{NN}}$=5.02 TeV, as well as for Xe-Xe at $\sqrt{s_{NN}}$=5.44 TeV, where the latter energies are almost twice as high. The small deviation evidenced in Xe-Xe collisions at low values of $\langle N_{part} \rangle$ could be due to the way in which the correlation between centrality and $\langle N_{part} \rangle$ is estimated in the standard Glauber MC approach \cite{Loizides17}. 
For consistency reasons, for Au-Au collisions ($\sqrt{s_{NN}}$ = 200 GeV) we have used both the data published by STAR and PHENIX Collaborations. As one can observe in Figure \ref{fig-4}a, there is very good agreement between these two datasets. In order to avoid overloaded figures as much as possible, from now on only the dataset measured by the STAR Collaboration will be used.

\begin{figure}
\centering
\includegraphics[scale=0.45]{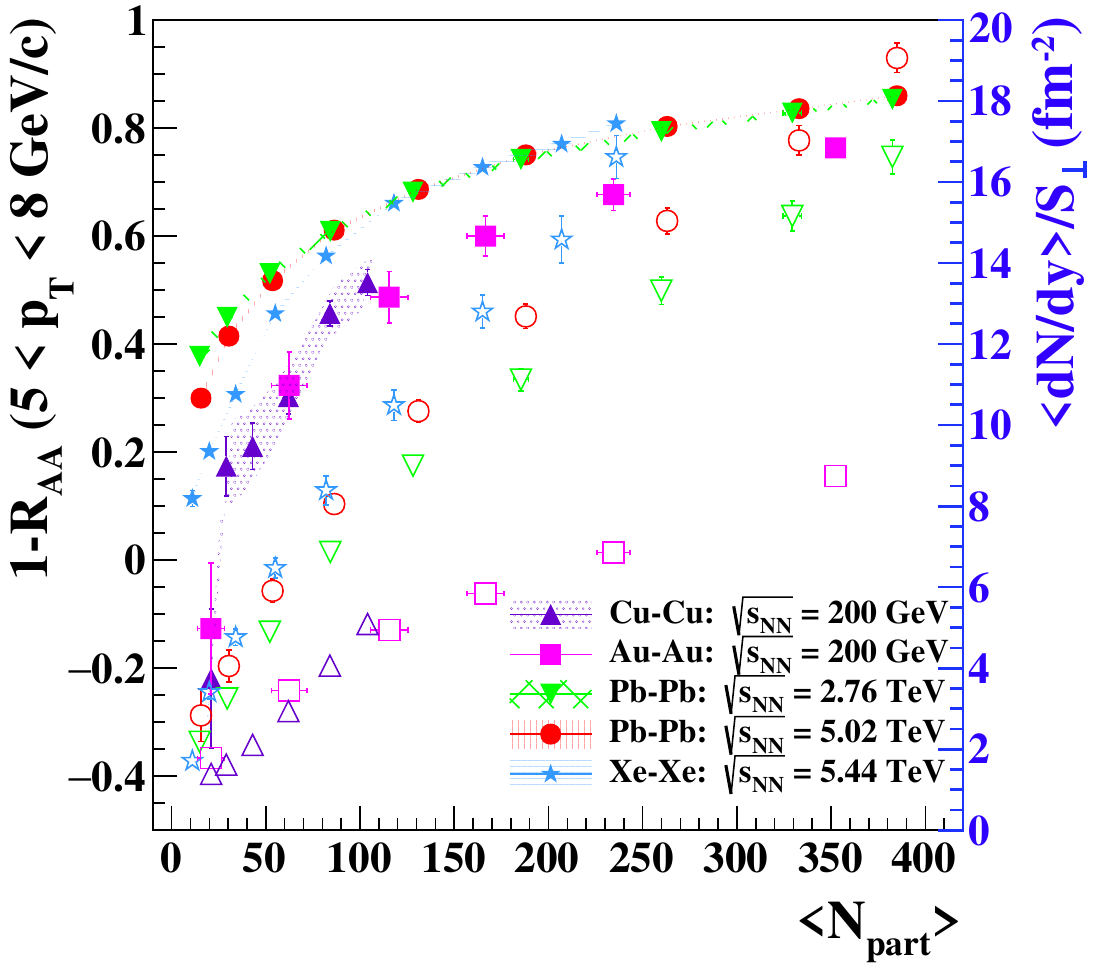}
\caption{(1-$R_{AA}$) (full symbols - left scale) and $\langle dN/dy \rangle/S_{\perp}$ (open symbols - right scale) as a function of $\langle N_{part} \rangle$. The abscissa is the same as in Figure \ref{fig-5} and consequently, at a given $\langle N_{part} \rangle$, the full and open symbols correspond to the same fireball shape.}
\label{fig-6}
\end{figure}
\begin{figure*}[t!]
\centering
\includegraphics[width=0.95\linewidth]{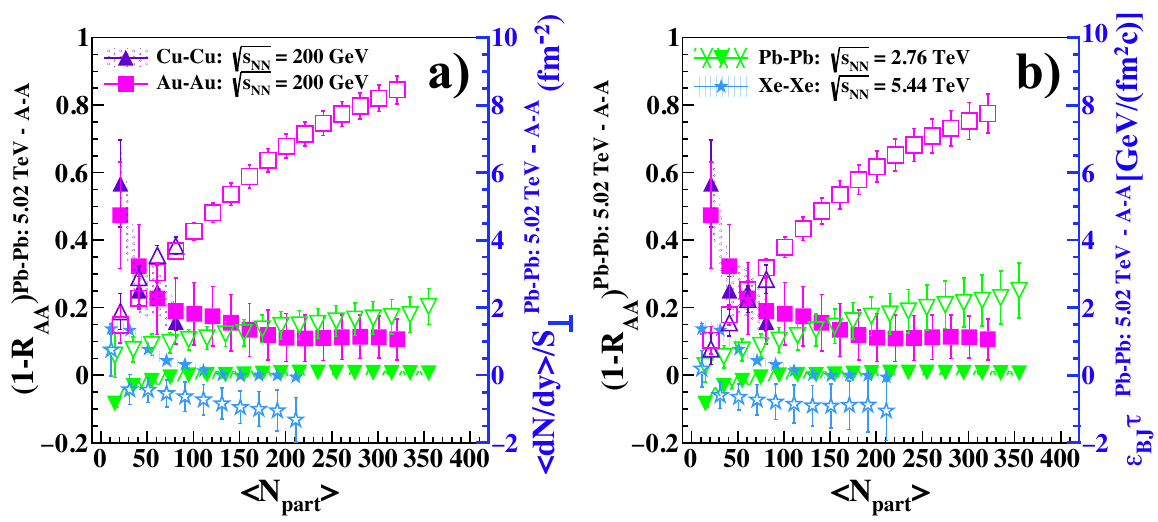}
\caption{The difference between the suppression in Pb-Pb at $\sqrt{s_{NN}}$=5.02 TeV and the suppression in Pb-Pb at $\sqrt{s_{NN}}$=2.76 TeV, Xe-Xe at $\sqrt{s_{NN}}$ = 5.44 TeV, Au-Au and Cu-Cu at $\sqrt{s_{NN}}$=200 GeV (full symbols). The corresponding differences in terms of particle density per unit of rapidity and unit of overlapping area $\langle dN/dy \rangle/S_{\perp}$ (Figure \ref{fig-7}a - open symbols) and of Bjorken energy density times the interaction time $\epsilon_{Bj}\cdot\tau$ (Figure \ref{fig-7}b - open symbols) at the corresponding collision energies can be followed using the scales on the right sides.}
\label{fig-7}
\end{figure*}

The  $\langle N_{part} \rangle$ dependence of the suppression has the advantage that at a given 
$\langle N_{part} \rangle$, the fireball transverse area $S_{\perp}$
is the same for the colliding systems and collision energies in question \cite{glauber_b}, with small deviations observed at very central collisions in Cu-Cu and Xe-Xe relative to Au-Au and Pb-Pb \cite{Petrovici18}, where the fireball shapes are closer to a circular geometry, qualitatively represented in Figure \ref{fig-5}. At LHC energies, with a slight change in the offset ($\approx$10 $fm^2$) the linear dependence of $S_{\perp}$ on $\langle N_{part} \rangle$ 
has the same slope as at the RHIC energy (Figure \ref{fig-3}).
 As it is known, all theoretical models predict a greater suppression with increasing path length and energy density or temperature of the deconfined medium traversed by a parton \cite{Baier97, Gyulassy00, Baier01, Arleo02, Muller05, Djordjevic08, Casalderrey-Solana15, Burke14, Betz14, Arleo18}.
 In Figure \ref{fig-6}, the suppression in terms of (1-$R_{AA}$) in the 5$<p_T<$8 GeV/c region for the colliding systems and energies under consideration, compared with the particle density per unit of rapidity and unit of overlapping area ($\langle dN/dy \rangle/S_{\perp}$), which is a measure of the entropy density and thus of the temperature \cite{Vogt}, as a function of $\langle N_{part} \rangle$, is represented. The $dN/dy$ values were estimated as in \cite{Petrovici18, PetAIP}. 
\begin{figure*}
\centering
\includegraphics[width=0.95\linewidth]{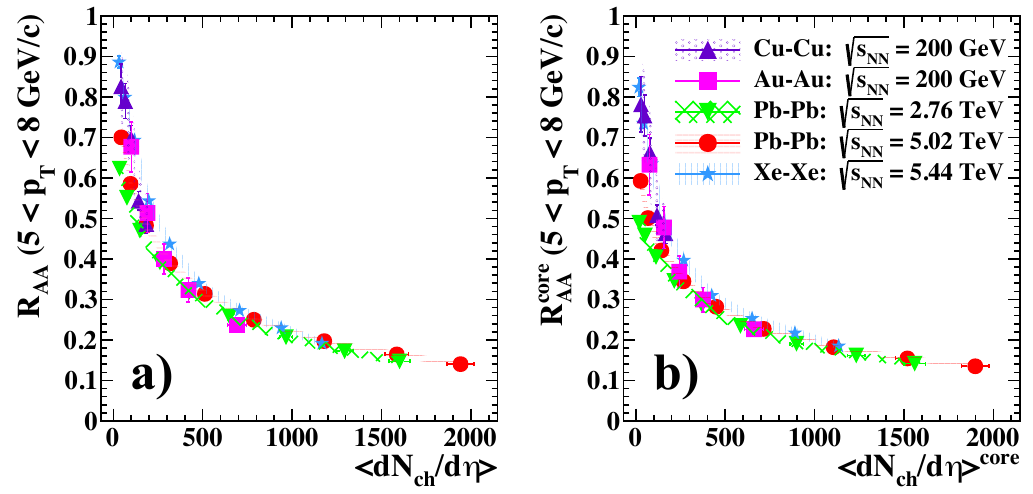}
\caption{$R_{AA}$ as a function of charged particle density per unit of pseudorapidity, $\langle dN_{ch}/d\eta \rangle$, for the same systems and collision energies as in Figure \ref{fig-4}; a) experimental values; b) the core contribution to $R_{AA}$ and $\langle dN_{ch}/d\eta \rangle$.}
\label{fig-8}
\end{figure*}

In the case of Cu-Cu and Au-Au at $\sqrt{s_{NN}}$=200 GeV, for the same average number of participants and $\langle dN/dy \rangle/S_{\perp}$, the suppression has the same value, increasing with $\langle dN/dy \rangle/S_{\perp}$ and with the size of the overlapping area. Since the suppression in central Cu-Cu collisions is the same as in Au-Au collisions at the corresponding $\langle N_{part} \rangle$, it appears that the fireball shape plays a minor role, for the same size of the overlapping area, on the azimuthally averaged $R_{AA}$ values. For $\langle N_{part} \rangle$=200, the differences in $\langle dN/dy \rangle/S_{\perp}$ for Pb-Pb at $\sqrt{s_{NN}}$=2.76, 5.02 TeV and for Xe-Xe at $\sqrt{s_{NN}}$=5.44 TeV, relative to Au-Au at $\sqrt{s_{NN}}$=200 GeV, are 5.25$\pm 1$, 6.77$\pm$1 and 7.89$\pm$1 (particles/$fm^2$) while the differences in (1-$R_{AA}$) are 0.10$\pm$0.03, 0.11$\pm$0.03 and 0.11$\pm$0.03. This suggests a suppression saturation at LHC energies. For central Au-Au collisions, i.e. $\langle N_{part} \rangle$=350, the difference in $\langle dN/dy \rangle/S_{\perp}$ between Pb-Pb at $\sqrt{s_{NN}}$=2.76 TeV and Au-Au at $\sqrt{s_{NN}}$=200 GeV is 
7$\pm$1 (particles/$fm^2$) while the difference in (1-$R_{AA}$) is 0.08$\pm$0.03. 
 
 Using the phenomenological "abc" parton energy loss approach \cite{Xu}, where the fractional energy loss is:
 \begin{equation}
 \frac{\Delta E}{E}\propto T^aL^b
 \end{equation}
 and the approximation from \cite{Djordjevic19}, one obtains:
 \begin{equation}
 (1-R_{AA})\propto \xi T^aL^b, 
 \end{equation}
where L is the average path length and T is the average temperature.
With the assumption that  
$L^2\propto S_{\perp}$, the entropy density $s  \propto \langle dN/dy \rangle /S_{\perp}$ and $T^3\propto \langle dN/dy \rangle/S_{\perp}$ \cite{Vogt}, 
one could estimate $\xi$ for three values of the parameters a and b, used by different models in order to reproduce the experimental results related to the suppression:
i) a=1, b=2; 
ii) a=1, b=1 \cite{Djordjevic19,zig20};
iii) a=3, b=2 \cite{Xu}.
 Using the experimental values of (1-$R_{AA}$) for Au-Au at $\sqrt{s_{NN}}$=200 
 GeV and Pb-Pb at $\sqrt{s_{NN}}$=2.76 TeV at $\langle N_{part} \rangle$=350, corresponding to the most central collisions for Au-Au at $\sqrt{s_{NN}}$=200 GeV, one obtains:
i) $\xi^{LHC}$ = 0.86($\pm 0.03$)$\cdot$$\xi^{RHIC}$; ii) $\xi^{LHC}$ = 0.88($\pm 0.03$)$\cdot$$\xi^{RHIC}$ and iii) $\xi^{LHC}$ = 0.58($\pm$ 0.04)$\cdot$$\xi^{RHIC}$.
Although theoretically not compelling, as it was mentioned in Ref. \cite{Betz14}, we used ansatz iii) based on their results on relative success and failure of different models (Table 2 and 3 in the same paper). Previous studies \cite{Betz13,Betz12,Betz11} have shown that the running coupling alters the jet-energy dependence of energy loss and $\frac{\Delta E}{E}$ is approximatively independent on E.  
Obviously, the hydrodynamic expansion of the deconfined matter traversed by the parton plays a role in the estimated final suppression. Using the $\sqrt{\langle dN/dy \rangle/S_{\perp}}$ scaling of the average transverse flow velocity, $\langle \beta_T \rangle$, reported in Ref. \cite{Petrovici18}, for the geometrical scaling variable corresponding to the particle densities used before for the $\xi^{LHC}/\xi ^{RHIC}$ estimation, a ratio 
$\langle \beta_T \rangle^{LHC}/\langle \beta_T \rangle^{RHIC}\simeq$1.09$\pm$0.08 is obtained. This could be one of the reasons leading to lower values of the jet-medium coupling in Pb-Pb central collision at $\sqrt{s_{NN}}$=2.76 TeV energy relative to Au-Au central collision at $\sqrt{s_{NN}}$=200 GeV. 
 In Table II $(\langle dN/dy \rangle/S_{\perp})^{1/3}$ $\sim$T, $\langle \beta_{T} \rangle$ and $(1-R_{AA})^{\pi^0}$ for the 0-5\% collision centrality for Au-Au at $\sqrt{s_{NN}}$=39 GeV and 200 GeV and for Pb-Pb at $\sqrt{s_{NN}}$=2.76 TeV are listed.  A comparison between 39 GeV and 200 GeV for Au-Au collisions shows an increase of 39.7\% relative to $\sqrt{s_{NN}}$=39 GeV in the suppression, while the increase in $(\langle dN/dy \rangle/S_{\perp})^{1/3}$ and $<\beta_T>$ is 20.3\% and 20.4\%, respectively. The increase in the suppression from $\sqrt{s_{NN}}$=200 GeV (Au-Au) to $\sqrt{s_{NN}}$=2.76 TeV is 7.4\% relative to $\sqrt{s_{NN}}$=200 GeV, while the increase in $(\langle dN/dy \rangle/S_{\perp})^{1/3}$ and $<\beta_T>$ is 24.2\% and 10.2\%, respectively. In this case, a $\approx$ 4 times smaller difference in the suppression, for a larger difference in $(\langle dN/dy \rangle/S_{\perp})^{1/3}$  and a smaller difference in the expansion velocity, can be observed.
\begin{longtable}{|c|c|c|c|c|c|}
\caption{$(\frac{\langle dN/dy \rangle}{S_{\perp}})^{1/3}$, $\langle \beta_{T} \rangle$ \cite{Petrovici18} and $(1-R_{AA})^{\pi^0}$ for the 0-5\% collision centrality for Au-Au at $\sqrt{s_{NN}}$=39 GeV and 200 GeV \cite{Adare12, Adare08} and for Pb-Pb at $\sqrt{s_{NN}}$=2.76 TeV
\cite{ALICE14}.}\\
\hline
\bf {System} & \bf $\sqrt{s_{NN}}$ & \bf Cen. & \bf $(\frac{\langle dN/dy \rangle/}{S_{\perp}})^{1/3}$ & \bf $\langle \beta_{T} \rangle$ &  \bf $(1-R_{AA})^{\pi^0}$\\ 
     & (GeV) & (\%)  &   &  & \\
\hline
\hline
Au-Au & 39 & 0-5 & 1.72$\pm$0.03 & 0.49$\pm$0.04 & 0.58$\pm$0.02\\
Au-Au & 200 & 0-5 & 2.07$\pm$0.03 & 0.59$\pm$0.05 & 0.81$\pm$0.06\\
Pb-Pb & 2760 & 0-5 & 2.57$\pm$0.04 & 0.65$\pm$0.06 & 0.87$\pm$0.08\\
\hline
\end{longtable} 
   This supports the assumption that the main contribution to the observed evolution of $(1-R_{AA})$ as a function of collision energy, i.e. a strong increase followed by a weakening dependence, is due to the energy density (temperature) dependence of the parton energy loss in the deconfined medium.  
 In Figure \ref{fig-7}, the relative differences between the suppression in Pb-Pb at $\sqrt{s_{NN}}$=5.02 TeV and the suppression in Au-Au and Cu-Cu collisions at $\sqrt{s_{NN}}$=200 GeV, Pb-Pb collisions at $\sqrt{s_{NN}}$=2.76 TeV and Xe-Xe collisions at 
$\sqrt{s_{NN}}$=5.44 TeV are shown. The corresponding differences in particle density per unit of rapidity and unit of overlapping area, $\langle dN/dy \rangle/S_{\perp}$ (Figure \ref{fig-7}a) and Bjorken energy density times the interaction time, $\epsilon_{Bj}\cdot\tau$ (Figure \ref{fig-7}b) are also shown with the corresponding scales on the right side of the figures. 

 The Bjorken energy density times the interaction time is estimated based on \cite{Bjorken82_2}:
 \begin{equation}
	\epsilon_{Bj}\cdot\tau = \frac{dE_{T}}{dy} \cdot \frac{1}{S_{\perp}}
\end{equation} 
 where $E_{T}$ is the total transverse energy and $S_{\perp}$ represents the overlapping area of the colliding nuclei.
 The total transverse energy per unit of rapidity can be estimated as follows: 
 \begin{itemize}
   \item RHIC $\sqrt{s_{NN}}$=200 GeV:
  	\begin{equation}
  	  \frac{dE_{T}}{dy} \approx\frac{3}{2}\left( \langle m_{T} \rangle\langle \frac{dN}{dy} \rangle \right)_{\pi^{\pm}} + 2 \left( \langle m_{T} \rangle\langle \frac{dN}{dy} \rangle\right)_{K^{\pm},p,\bar{p}} 
    \end{equation}
   \item LHC energies:
    \begin{equation}
    	\begin{aligned}
  	  \frac{dE_{T}}{dy} &\approx\frac{3}{2}\left(\langle m_{T} \rangle\langle \frac{dN}{dy} \rangle \right)_{\pi^{\pm}} + 2\left(\langle m_{T} \rangle\langle \frac{dN}{dy} \rangle\right)_{K^{\pm},p,\bar{p},\Xi^{-},\bar{\Xi}^{+}} \\ 
  	  					&+ \left(\langle m_{T} \rangle\langle \frac{dN}{dy} \rangle\right)_{\Lambda, \bar{\Lambda},\Omega^{-},\bar{\Omega}^{+}}
  	 \end{aligned}
    \end{equation}
 \end{itemize}
The input data used in the estimation of the Bjorken energy density times the interaction time are reported in \cite{Petrovici18, STAR_dNchdeta, PbPb_276_dNchdeta, Jacazio17, Adams07, Abelev13_2, Albuquerque18, Brahms16, Bellini19}  and  Table I.
 
Within the error bars, the suppression in Pb-Pb collisions at $\sqrt{s_{NN}}$=2.76 TeV is the same with the one corresponding to $\sqrt{s_{NN}}$=5.02 TeV for all values of $\langle N_{part} \rangle$, although the difference in $\langle dN/dy \rangle/S_{\perp}$ and in $\epsilon_{Bj} \cdot \tau$ increases from 0.88$\pm$0.33 particles/$fm^2$ to 1.95$\pm$0.54 particles/$fm^2$ and from 0.71$\pm$0.32 GeV/($fm^2c$) to 2.44$\pm$0.81 GeV/($fm^2c$), respectively, from the low ($\langle N_{part} \rangle$=50) to the highest value of $\langle N_{part} \rangle$ ($\langle N_{part} \rangle$=350). The difference between the suppression in Pb-Pb at $\sqrt{s_{NN}}$=5.02 TeV and Au-Au at $\sqrt{s_{NN}}$=200 GeV decreases from 0.27$\pm$0.25 to 0.08$\pm 0.02$ with $\langle N_{part} \rangle$, while the differences in $\langle dN/dy \rangle/S_{\perp}$ and $\epsilon_{Bj} \cdot \tau$ increase from 2.63$\pm$0.29 particles/$fm^2$ and 2.13$\pm$0.28 GeV/($fm^2c$) to 8.9$\pm$0.43 particles/$fm^2$ and 8.2$\pm$0.8 GeV/($fm^2c$), respectively.

\begin{figure}
\centering
\includegraphics[scale=0.48]{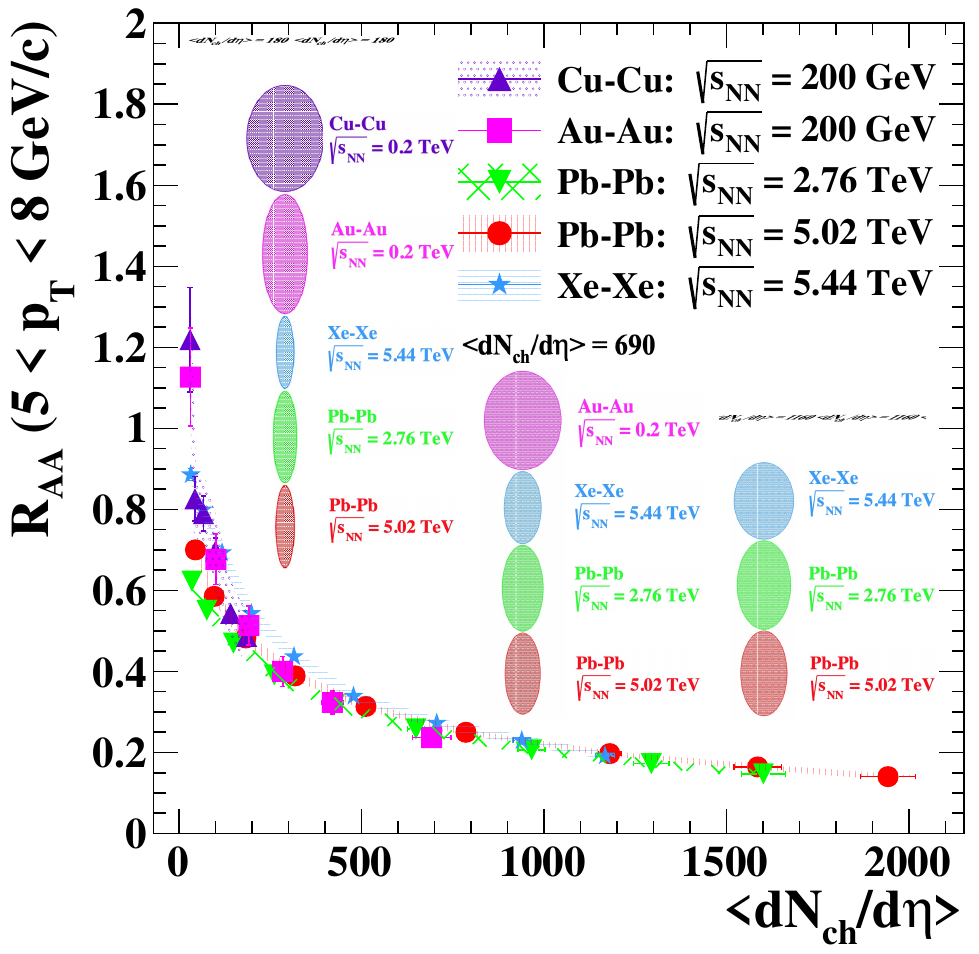}
\caption{The same as Figure \ref{fig-8}a, with the shapes of the overlapping area $S_\perp$ at different values of $\langle dN_{ch}/d\eta \rangle$.} 
\label{fig-9}
\end{figure}
\begin{figure}
\centering
\includegraphics[scale=0.48]{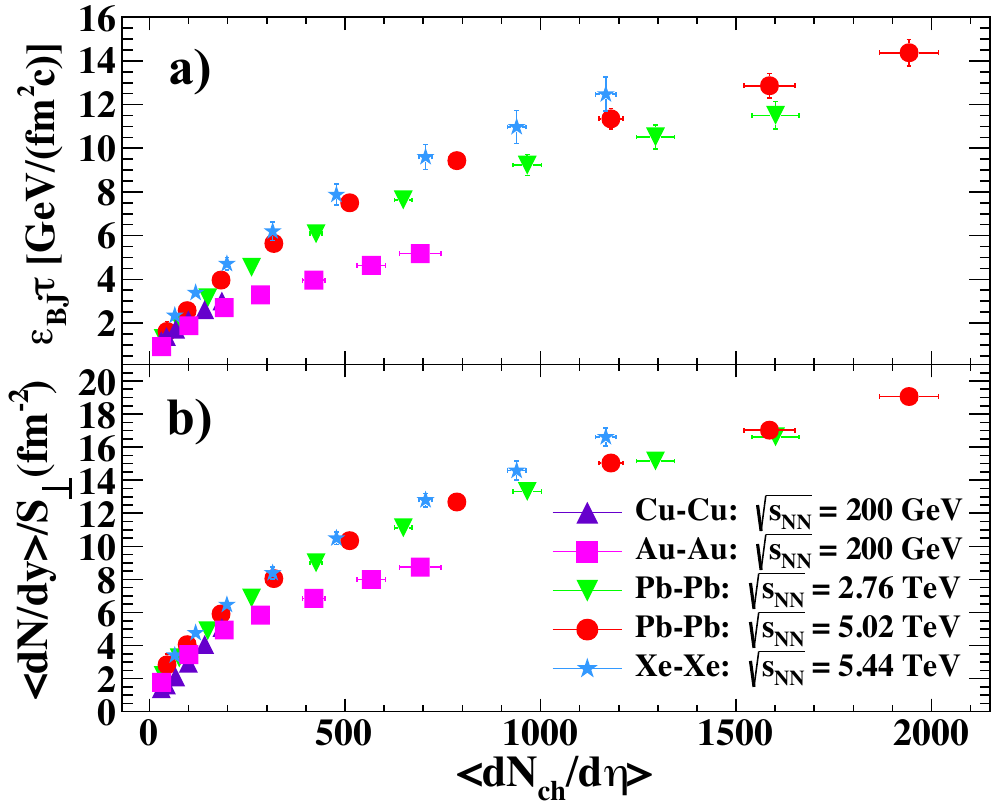}
\caption{a) The Bjorken energy density times the interaction time $\epsilon_{Bj}\cdot\tau$; b) particle density per unit of rapidity and unit of overlapping area $\langle dN/dy \rangle/S_{\perp}$, both as a function of the average charged particle density per unit of pseudorapidity $\langle dN_{ch}/d\eta \rangle$.}
\label{fig-10}
\end{figure} 
\begin{figure}
    \centering
    \includegraphics[width=0.95\linewidth]{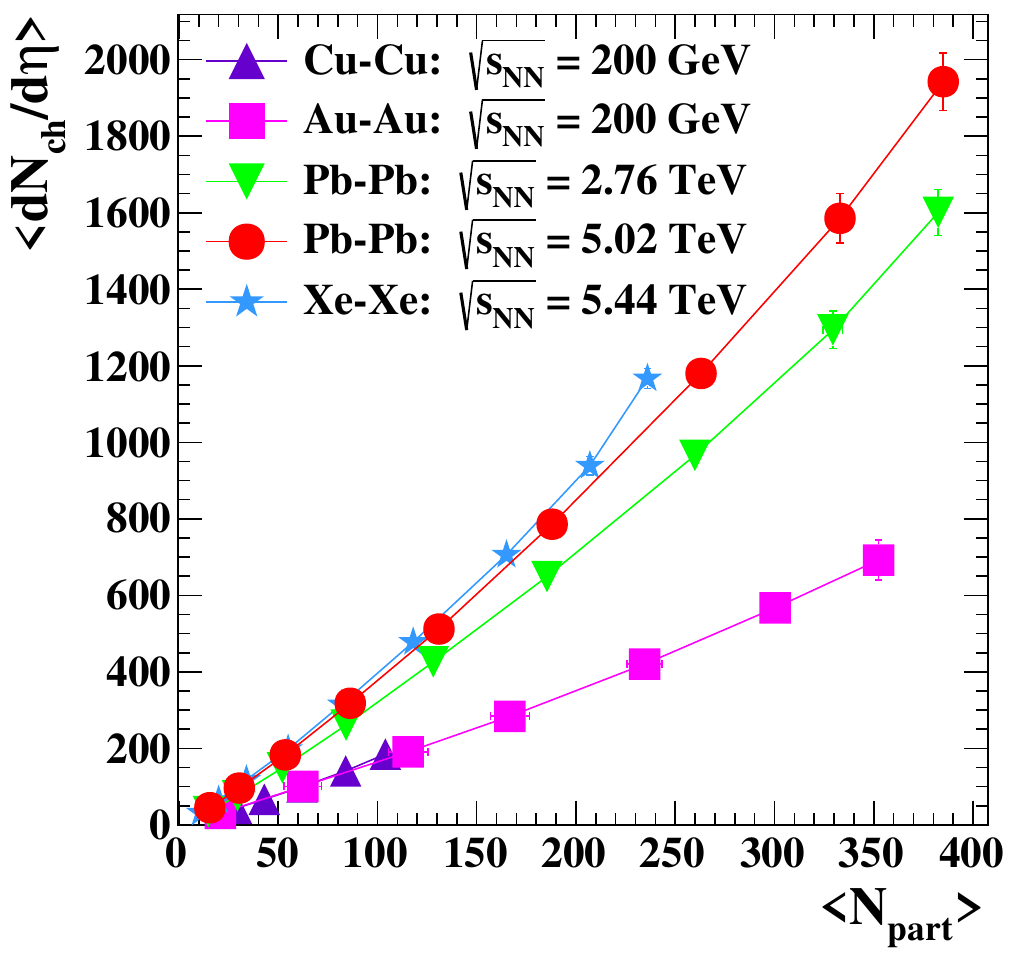}[h!]
\caption{The average charged particle density per unit of pseudorapidity $\langle dN_{ch}/d\eta \rangle$ as a function of the average value of participating nucleons $\langle N_{part} \rangle$.}
\label{fig-11}
\end{figure}
 An alternative representation of $R_{AA}$ could be done as a function of the average charged particle density per unit of pseudorapidity \cite{XeXe_data}. The 
$\langle dN_{ch}/d \eta \rangle$ experimental data for heavy ion collisions are taken from 
\cite{XeXe_data, STAR_dNchdeta, BES_CuCu_dNchdeta, PbPb_502_dNchdeta, PbPb_276_dNchdeta}. 
The $R_{AA}$ as a function of $\langle dN_{ch}/d \eta \rangle$ is presented in Figure \ref{fig-8}a for the same systems and collision energies as in Figure \ref{fig-4}. In such a representation, all systems at all energies scale as a function of $\langle dN_{ch}/d\eta \rangle$. The same representation in terms of $R_{AA}^{core}$ and $\langle dN_{ch}/d\eta \rangle^{core}$ (Figure \ref{fig-8}b) shows a larger deviation between RHIC and LHC energies for 
 $\langle dN_{ch}/d\eta \rangle \leq$ 200. Relative to the 
 $\langle N_{part} \rangle$ dependence, the difference in the shapes of the overlapping areas of different systems for a given $\langle dN_{ch}/d\eta \rangle$ is larger, as it can be seen in Figure \ref{fig-9}. 
 If we look at the dependence of $\epsilon_{Bj}\cdot\tau$ or $\langle dN/dy \rangle/S_{\perp}$ respectively as a function of charged particle density (Figure \ref{fig-10}a and Figure \ref{fig-10}b), a difference between the collision energies which increases with $\langle dN_{ch}/d\eta \rangle$ is seen.
  
 Therefore, with several contributions playing a role in the observed scaling in $\langle dN_{ch}/d\eta \rangle$, it is rather difficult to unravel the importance of each one of them. The difference between the two representations is explained by the correlation between $\langle dN_{ch}/d\eta \rangle$ and $\langle N_{part} \rangle$, presented in Figure \ref{fig-11}. 
 While the overlapping area depends little on the system size and collision energy for a given $\langle N_{part} \rangle$ \cite{Petrovici18}, $\langle dN_{ch}/d\eta \rangle$ combines the contribution of both collision energy and system size.

\section{Why $R_{AA}^N$ ?}   

\begin{figure}
    \centering
\includegraphics[width=0.95\linewidth]{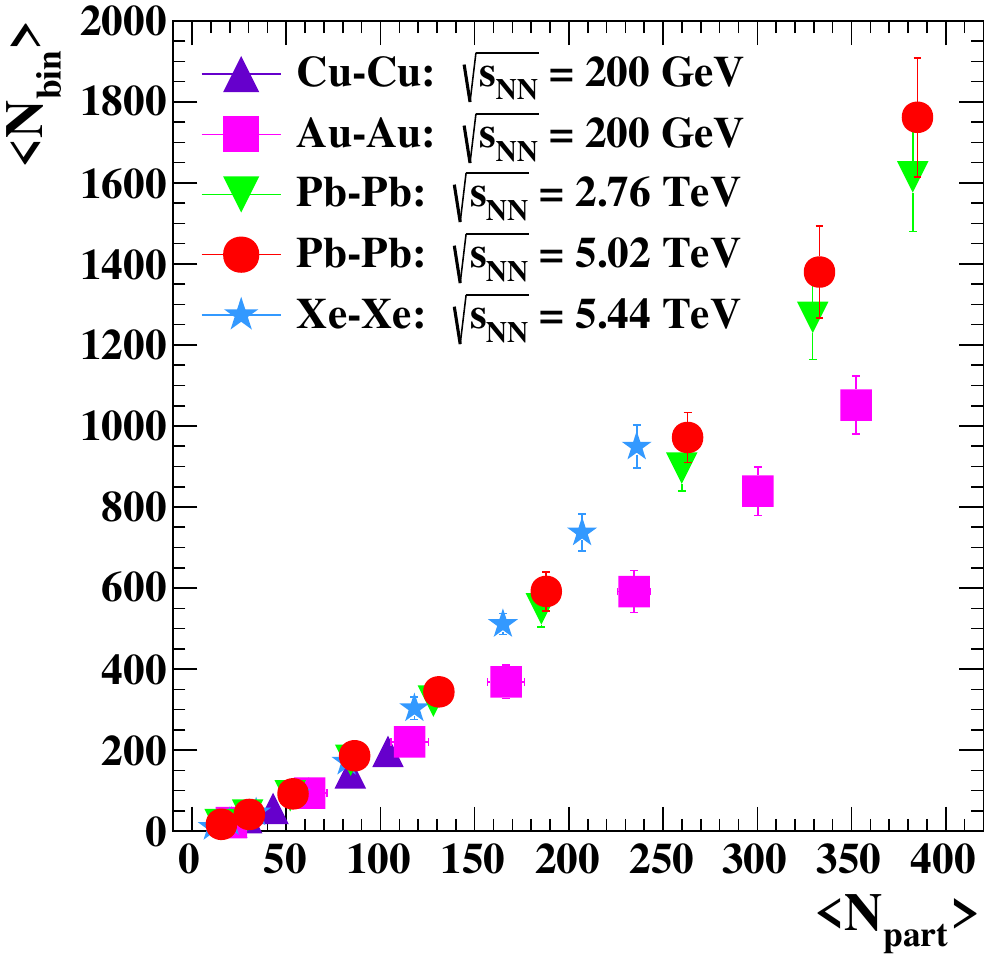}
\caption{Correlation between the average number of binary collisions $\langle N_{bin} \rangle$ and the average number of participating nucleons $\langle N_{part} \rangle$ estimated using the Glauber MC approach.}
\label{fig-12}
\end{figure}

 $R_{AA}$, as a measure of the suppression in heavy ion collisions, is based on the estimate of the number of binary collisions $\langle N_{bin} \rangle$ within the Glauber MC approach using  straight trajectories as a hypothesis. The dependence on the collision energy is introduced by the nucleon-nucleon cross section and the oversimplified assumption that every nucleon-nucleon collision takes place at the same energy, $\sqrt{s}$, and consequently the same cross section, $\sigma_{pp}$. 
 In Figure \ref{fig-12}, the correlation between the number of binary collisions $\langle N_{bin} \rangle$ and $\langle N_{part} \rangle$ estimated within the standard Glauber MC approach is represented.
  
 An alternative approach, where the energy and $\sigma_{pp}$ change after each collision \cite{Seryakov16}, has shown that in Pb-Pb collisions at $\sqrt{s_{NN}}$=2.76 TeV, the average number of binary collisions $\langle N_{bin} \rangle$ is significantly lower than the values estimated by the standard Glauber model with the difference increasing towards central collisions. The difference in $\langle N_{part} \rangle$ is negligible at peripheral and central collisions. For mid-central collisions it is about 18\%.  

\begin{figure}
    \centering
\includegraphics[width=\linewidth]{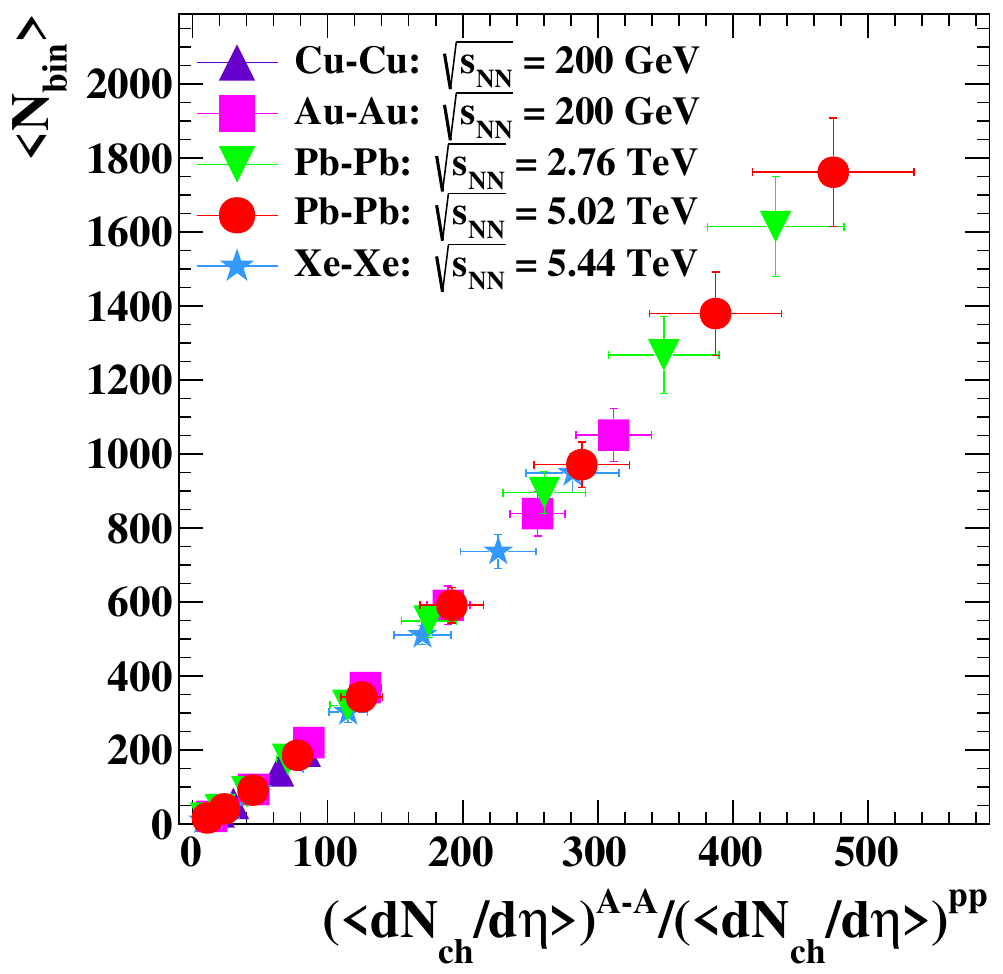}
\caption{Correlation between the average number of binary collisions $\langle N_{bin} \rangle$ and experimental $\langle dN_{ch}/d\eta \rangle^{AA}/\langle dN_{ch}/d\eta \rangle^{pp}$.}
\label{fig-13}
\end{figure}

\begin{figure}[t]
    \centering
    \includegraphics[width=\linewidth]{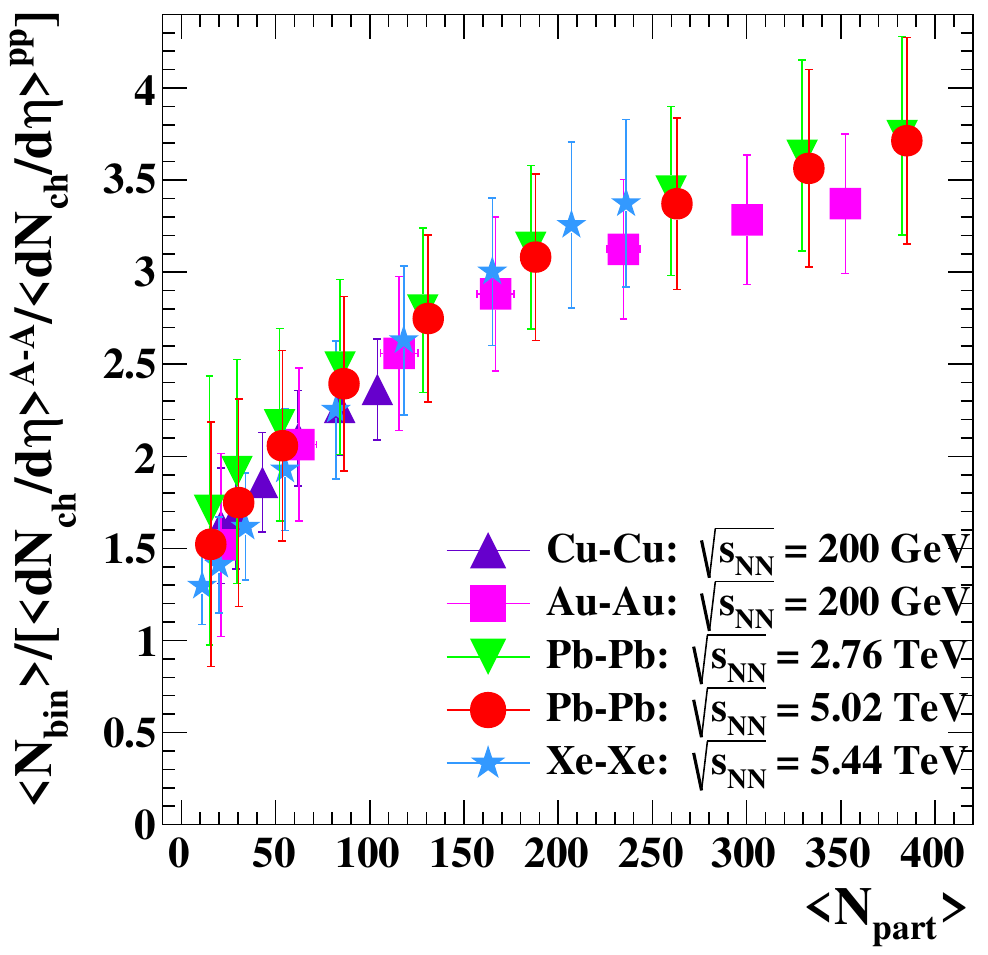}
\caption{$\langle N_{bin} \rangle/[\langle dN_{ch}/d\eta \rangle^{A-A}/\langle dN_{ch}/d\eta \rangle^{pp}]$ as a function of $\langle N_{part} \rangle$.}
\label{fig-14}
  \end{figure}

$\langle N_{bin} \rangle/[\langle dN_{ch}/d\eta \rangle^{A-A}/\langle dN_{ch}/d\eta \rangle^{pp,MB}]$ has to be unity if only single collisions take place. A very good correlation between $\langle N_{bin} \rangle$ estimated within the standard  Glauber model and experimental values of $\langle dN_{ch}/d\eta \rangle^{A-A}/\langle dN_{ch}/d\eta \rangle^{pp,MB}$ is evidenced in Figure \ref{fig-13}. 
 However, their ratio as a function of $\langle N_{part} \rangle$ shows an increase from close to 1 for the lowest values of $\langle N_{part} \rangle$, up to $\langle N_{part} \rangle$$\approx$150, followed by a tendency towards a saturation at $\approx$3.5 for the largest $\langle N_{part} \rangle$ values (Figure \ref{fig-14}). All systems at all investigated energies overlap in this representation. In the case of pp collisions, $\langle dN_{ch}/d \eta \rangle^{INEL}$ corresponding to the selection of inelastic collisions and the parametrisation given in \cite{ALICE17} had been used.
  \begin{figure}
    \centering
\includegraphics[width=\linewidth]{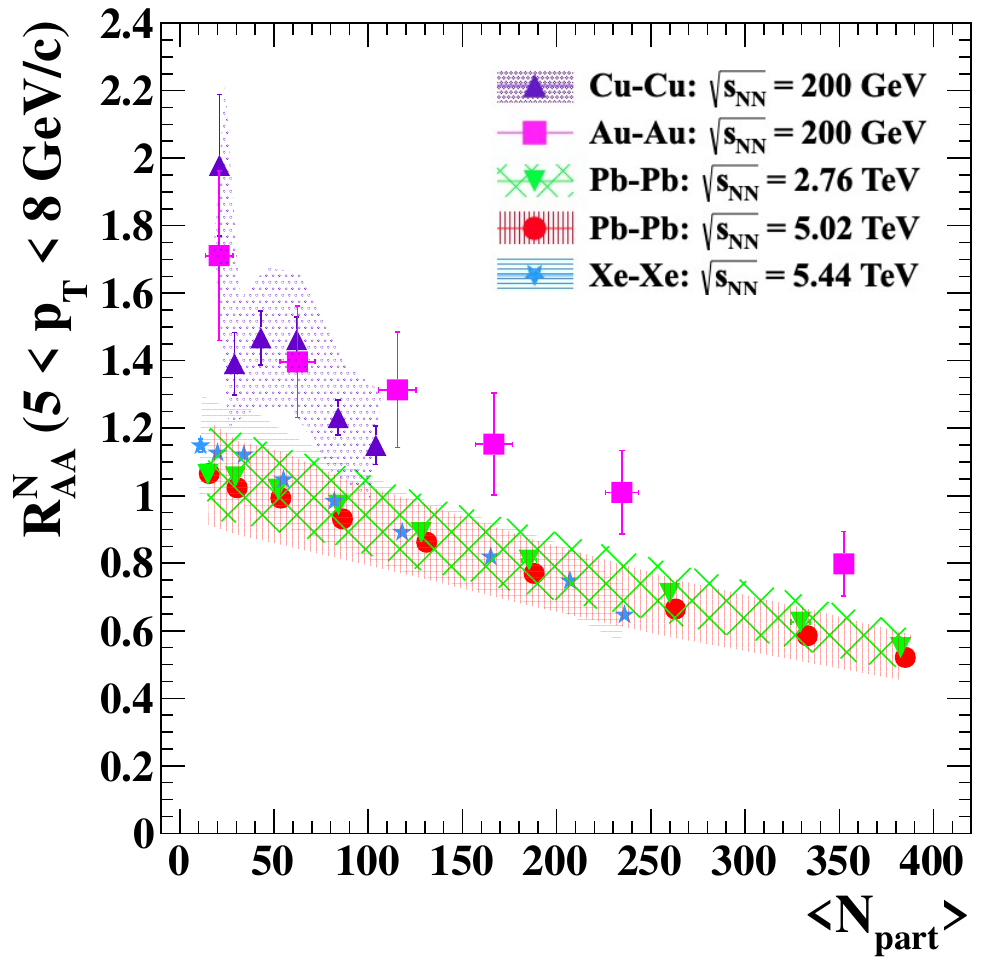}
\caption{$R_{AA}^N$ as a function of $\langle N_{part} \rangle$.}
\label{fig-15}
\end{figure}
 
 Based on these, we will also analyse the model independent quantity, namely $R_{AA}^N$, obtained as a ratio of the $p_T$ spectra in A-A collisions to the one in MB pp collisions at the same energy, with each of them normalised to the corresponding charged particle densities, for all the available centralities in A-A collisions (Eq.(11)). This observable was used in a previous paper for comparing the behaviour of $p_T$ spectra in pp, p-Pb and Pb-Pb collisions as a function of charged particle multiplicity and centrality, respectively \cite{Petrovici17_aip}. 
\begin{equation}
R_{AA}^N=\frac{(\frac{d^{2}N}{dp_{T}d\eta}/\langle \frac{dN_{ch}}{d\eta}\rangle)^{cen}}{(\frac{d^{2}N}{dp_{T}d\eta}/\langle\frac{dN_{ch}}{d\eta}\rangle)^{pp,MB}}
\end{equation}   

\begin{figure}[t]
    \centering
    \includegraphics[width=\linewidth]{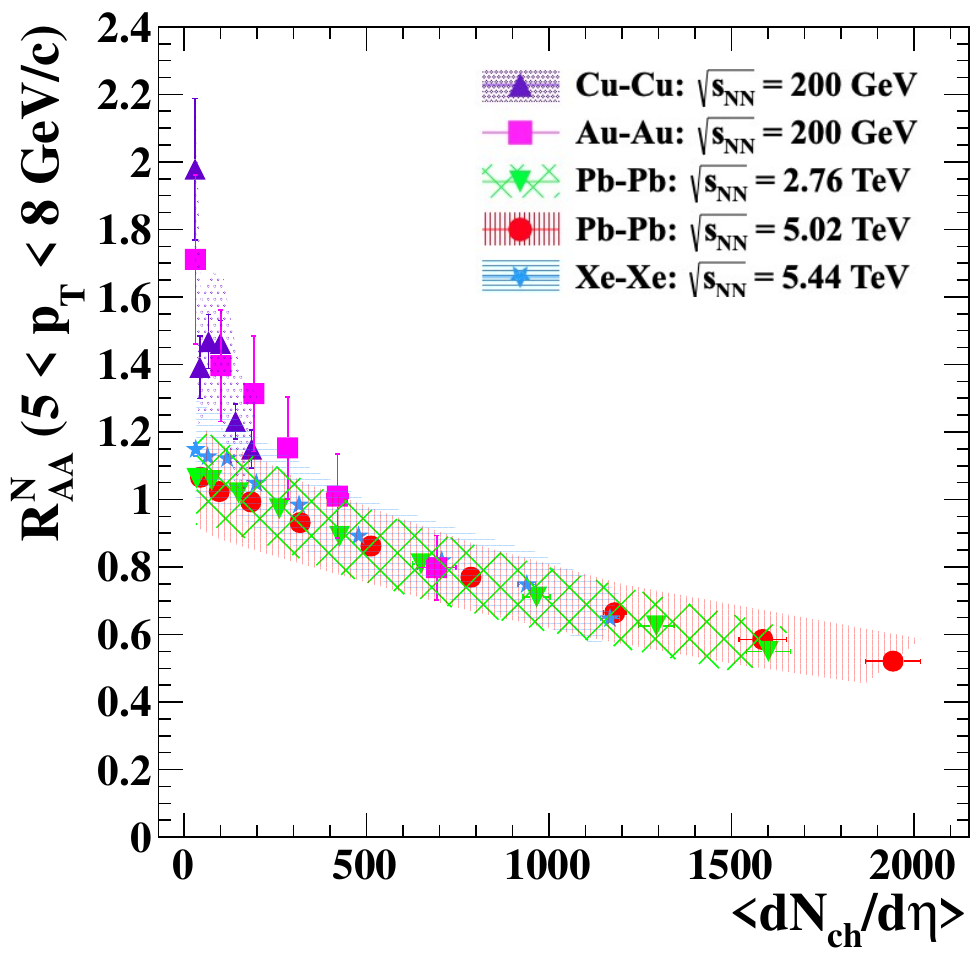}
\caption{$R_{AA}^{N}$ as a function of $\langle dN_{ch}/d\eta \rangle$.}
\label{fig-16}
  \end{figure}

 In Figure \ref{fig-15}, $R_{AA}^N$ as a function of $\langle N_{part} \rangle$ for the systems discussed in the previous section is presented. 
 The system size scaling for each energy domain, i.e. the highest energy at RHIC and LHC energies remains. $R_{AA}^N$ has a close to linear dependence  as a function of  
 $\langle N_{part} \rangle$ and at larger values of the average number of participating nucleons, the suppression is reduced compared to $R_{AA}$. As it is observed in Figure \ref{fig-16}, the scaling of $R_{AA}^N$ has not the same quality as $R_{AA}$ as a function of $\langle dN_{ch}/d\eta \rangle$ (see Figure \ref{fig-8}a) for the two collision energy domains. However, the scaling at LHC energies remains, a close to linear dependence being evidenced in this representation as well. 
 The same considerations as in Section III can be used in order to estimate the expected suppression, (1-$R_{pp}^{N(HM)}$), for pp collisions at $\sqrt{s}$=7 TeV and very high charged particle multiplicity (HM) events. The geometrical scaling \cite{Petrovici18} shows that for the highest charged particle multiplicity in pp collisions at $\sqrt{s}$=7 TeV, in the case of $\alpha$=1, $\sqrt{\langle dN/dy \rangle/S_{\perp}}$=3.3$\pm$0.1 particles$\cdot$$fm^{-1}$, $\langle \beta_T \rangle$ in pp and Pb-Pb at $\sqrt{s_{NN}}$=2.76 TeV is the same. Therefore, the contribution of the hydrodynamic expansion to the suppression should play a similar role. For this value of $\sqrt{\langle dN/dy \rangle/S_{\perp}}$, $S_{\perp}^{pp}$($\alpha$=1)=7.43$\pm$0.48 $fm^2$ and $S_{\perp}^{Pb-Pb}$=70$\pm$0.4 $fm^2$ (corresponding to $\langle N_{part}\rangle$=125). Assuming the same jet-medium coupling, 
 (1-$R_{pp}^{N(HM)}$)/(1-$R_{AA}^N({\langle N_{part} \rangle=125)}$) $\approx$ $S_{\perp}^{pp,HM}$/$S_{\perp}^{Pb-Pb, \langle N_{part} \rangle=125}$=0.11 $\pm$0.01. This could explain why in pp collisions at LHC, in high charged particle multiplicity events, in the limit of current experimental uncertainties, no suppression was observed, although similarities to Pb-Pb collisions for other observables were evidenced.

\section{Relative suppression in terms of $R_{CP}$}
 For energies where the $p_T$ spectra in pp collisions were not measured, the suppression was studied in terms of $R_{CP}$, i.e. the ratio of charged particle $p_T$ spectra at a given centrality to the $p_T$ spectrum in peripheral collisions, each of them divided by the corresponding average number of the binary collisions:
\begin{equation}
R_{CP}=\left(\frac{\frac{d^2N}{d\eta dp_T}}{\langle N_{bin}\rangle}\right)^{cen} / \left(\frac{\frac{d^2N}{d\eta dp_T}}{\langle N_{bin} \rangle}\right)^{peripheral}
\end{equation}
for each centrality in A-A collisions.
  \begin{figure}
    \centering
\includegraphics[width=\linewidth]{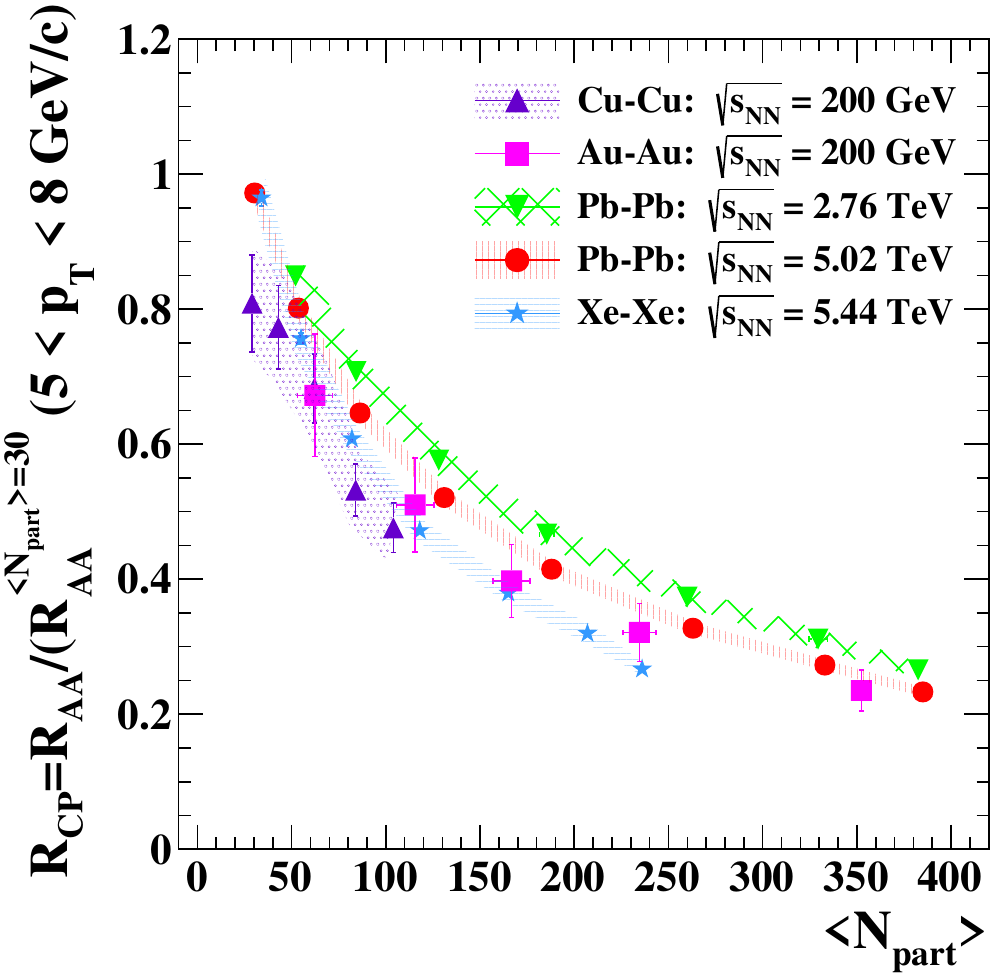}
\caption{$R_{CP}$ for Au-Au and Cu-Cu at 
$\sqrt{s_{NN}}$=200 GeV, Pb-Pb at $\sqrt{s_{NN}}$=2.76 TeV and 5.02 TeV and Xe-Xe at $\sqrt{s_{NN}}$=5.44 TeV, as a function of $\langle N_{part} \rangle$.}
\label{fig-17}
\end{figure}

 For a better comparison of the $R_{CP}$ values as a function of $\langle N_{part} \rangle$, the  peripheral collision of reference was chosen to be the same for all systems and all energies, i.e $\langle N_{part}\rangle$=30. The $R_{CP}$ estimated in this way is represented in Figure \ref{fig-17} for the same systems and energies. 
As in the case of $R_{AA}$, due to the same reasons, using experimental data, we estimated the $R_{CP}^N$:
\begin{equation}
R_{CP}^N=\left(\frac{\frac{d^2N}{d\eta dp_T}}{\langle \frac{dN_{ch}}{d\eta}\rangle}\right)^{cen} /\left(\frac{\frac{d^2N}{d\eta dp_T}}{\langle \frac{dN_{ch}}{d\eta} \rangle}\right)^{peripheral}
\end{equation}
\begin{figure}[h]
    \centering
    \includegraphics[width=\linewidth]{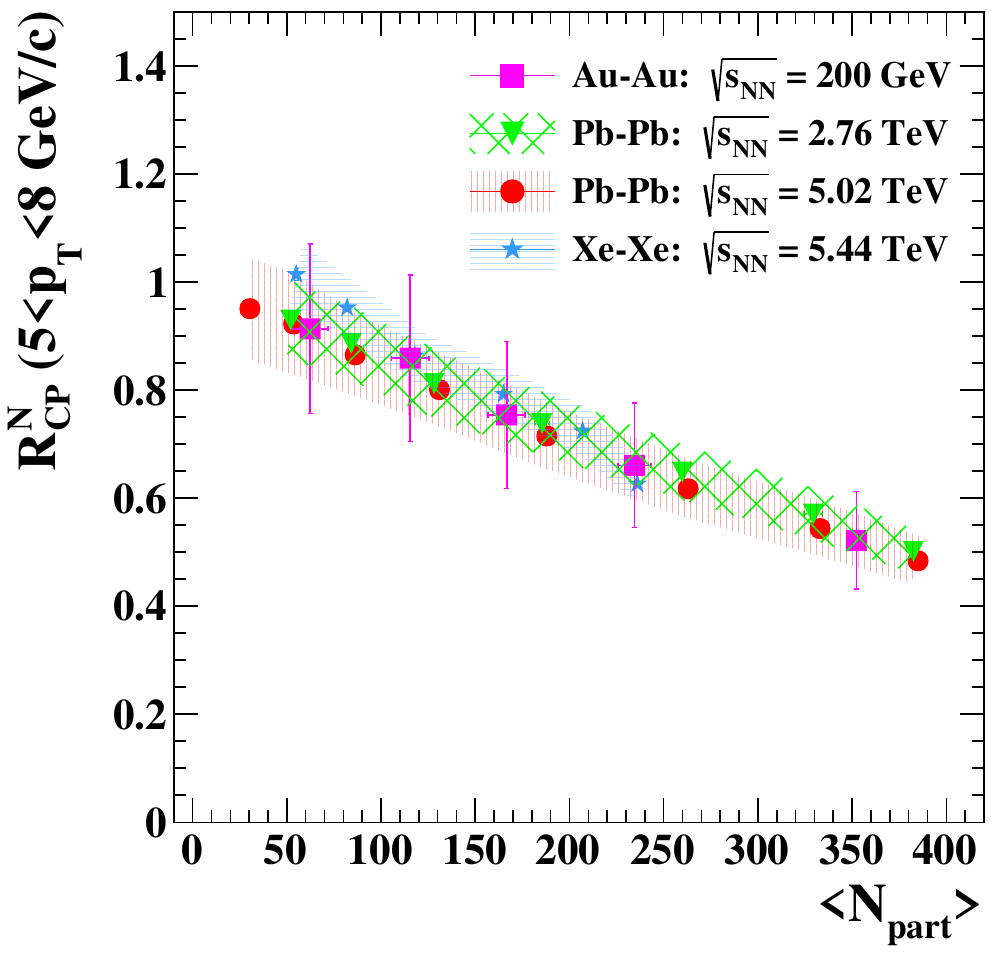}
\caption{$R_{CP}^N$ for Au-Au at $\sqrt{s_{NN}}$=200 GeV, Xe-Xe at $\sqrt{s_{NN}}$=5.44 TeV and Pb-Pb at $\sqrt{s_{NN}}$=2.76 TeV and 5.02 TeV as a function of $\langle N_{part} \rangle$.}
\label{fig-18}
  \end{figure}

 The $R_{CP}^N$ suppression as a function of $\langle N_{part} \rangle$ (Figure \ref{fig-18}) is the same at all values of $\langle N_{part} \rangle$ for all the heavy systems, Au-Au, Xe-Xe and Pb-Pb, although the difference in the collision energies is $\approx$ 14-27 times higher at LHC than at RHIC
and between 
the LHC energies is a factor of $\approx$ 2. The linear dependence as a function of $\langle N_{part} \rangle$ follows the linear dependence observed in $R_{AA}^N$.

\section{(1-$R_{AA}$)/$\langle dN/dy \rangle$ and (1-$R_{AA}^N$)/$\langle dN/dy \rangle$ dependence on $(\langle dN/dy \rangle/S_{\perp})^{1/3}$}

 Based on Eq. (7) and ansatz (iii) from Section III and taking $ S_{\perp} \propto L^2$, $\xi$, which is a rough estimate of the jet-coupling constant, turns out to be proportional to (1-$R_{AA}$)/ $\langle dN/dy \rangle$. 
 A qualitative temperature dependence of $\xi$ can be obtained from experimental data, as 
 \newline
 \noindent
 $T\sim(<dN/dy>/S_{\perp})^{1/3}$. 
  \begin{figure}[h]
    \centering
\includegraphics[width=\linewidth]{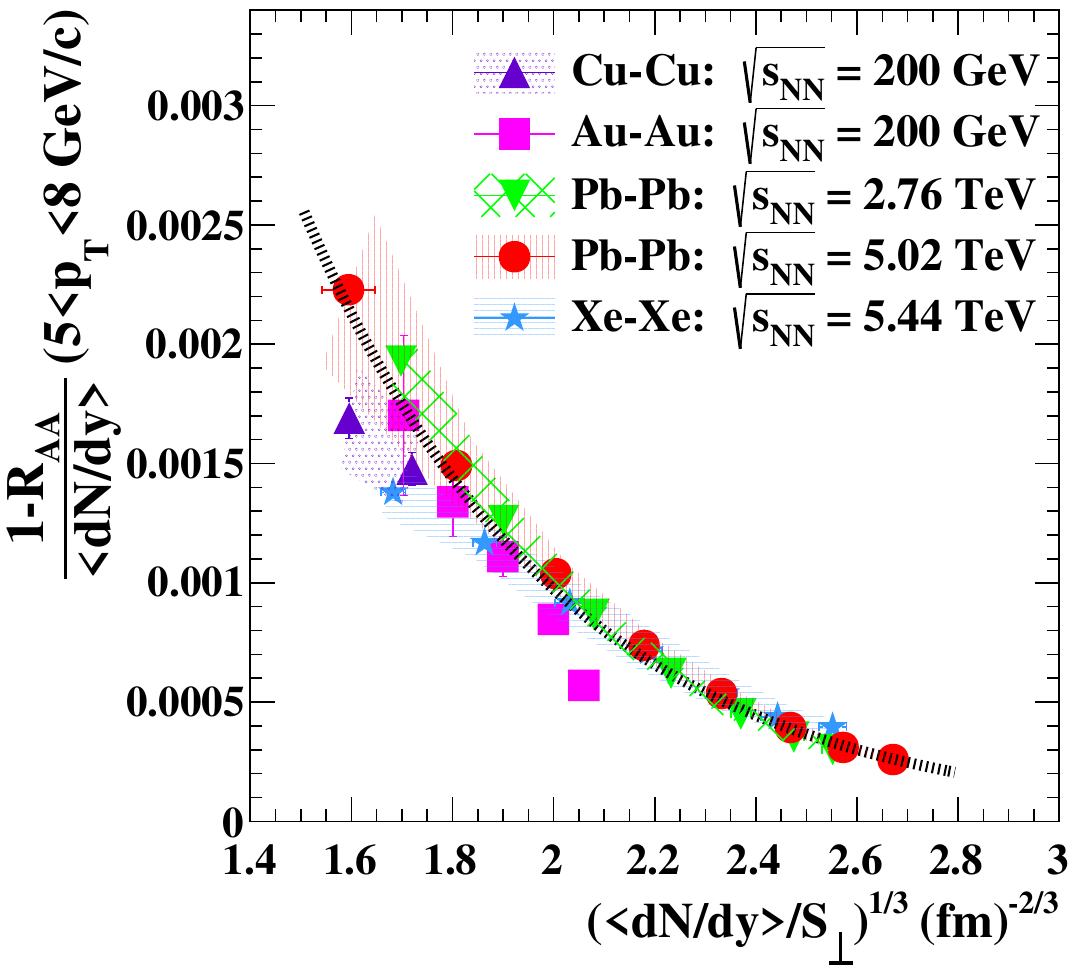}
\caption{(1-$R_{AA}$)/ $\langle dN/dy \rangle$ dependence on $(\langle dN/dy \rangle/S_{\perp})^{1/3}$. The line is the result of the fit with the expression (14).}
\label{fig-19}
\end{figure}
 \begin{figure}
\centering
\includegraphics[scale=0.48]{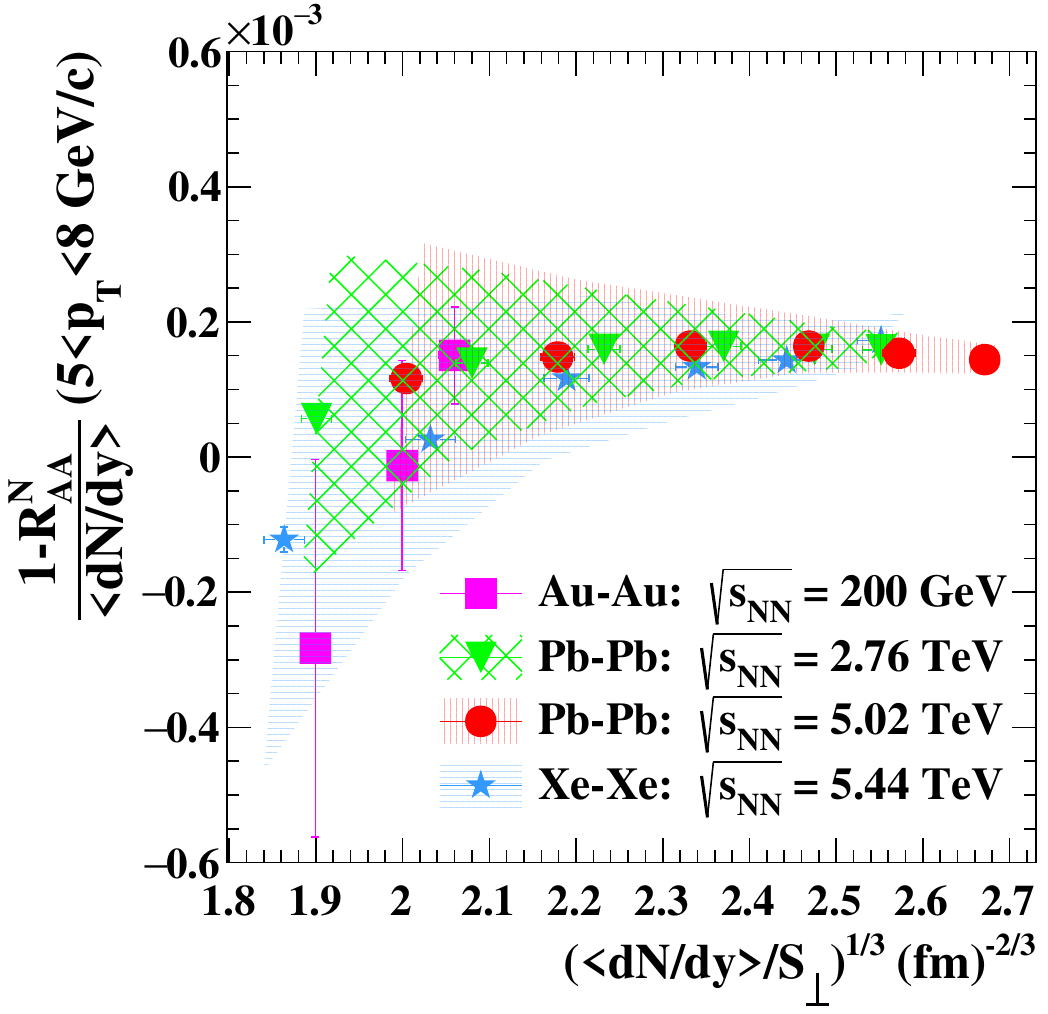}
\caption{(1-$R_{AA}^N$)/$\langle dN/dy \rangle$ dependence on $(\langle dN/dy \rangle/S_{\perp})^{1/3}$.}
\label{fig-20}
\end{figure} 
 
 As can be seen in Figure \ref{fig-19}, 
(1-$R_{AA}$)/$\langle dN/dy \rangle $ shows an exponential decrease as a function of $(\langle dN/dy \rangle/S_{\perp})^{1/3}$. The hatched line is the result of the fit with the following expression:
\begin{equation}
\frac{1-R_{AA}}{\langle dN/dy \rangle} = e^{\alpha -\beta\cdot (\langle dN/dy \rangle/S_{\perp})^{1/3}}
\end{equation}
Such a temperature dependence of the jet-medium coupling was considered  in Ref.\cite{Betz14} in
order to reproduce the nuclear modification factors at RHIC and LHC energies.
 A similar representation for $R_{AA}^N$ instead of $R_{AA}$ is presented in Figure \ref{fig-20}. In this case, (1-$R_{AA}^N$)/$\langle dN/dy \rangle$ is constant as a function of $(\langle dN/dy \rangle/S_{\perp})^{1/3}$, for $(\langle dN/dy \rangle/S_{\perp})^{1/3} \geq$ 2.1 
particles/fm$^{2/3}$, independent on the size of the colliding systems and collision energy. An impact parameter independence of the jet quenching parameter was
claimed in a series of theoretical estimates \cite{Andres, Andres2, Xie}.

\section{ The $\sqrt{s_{NN}}$ dependence of $R_{CP}$, $R_{CP}^N$, $R_{AA}^{\pi^0}$, $(R_{AA}^N)^{\pi^0}$} 

\begin{figure}
\centering
\includegraphics[scale=0.5]{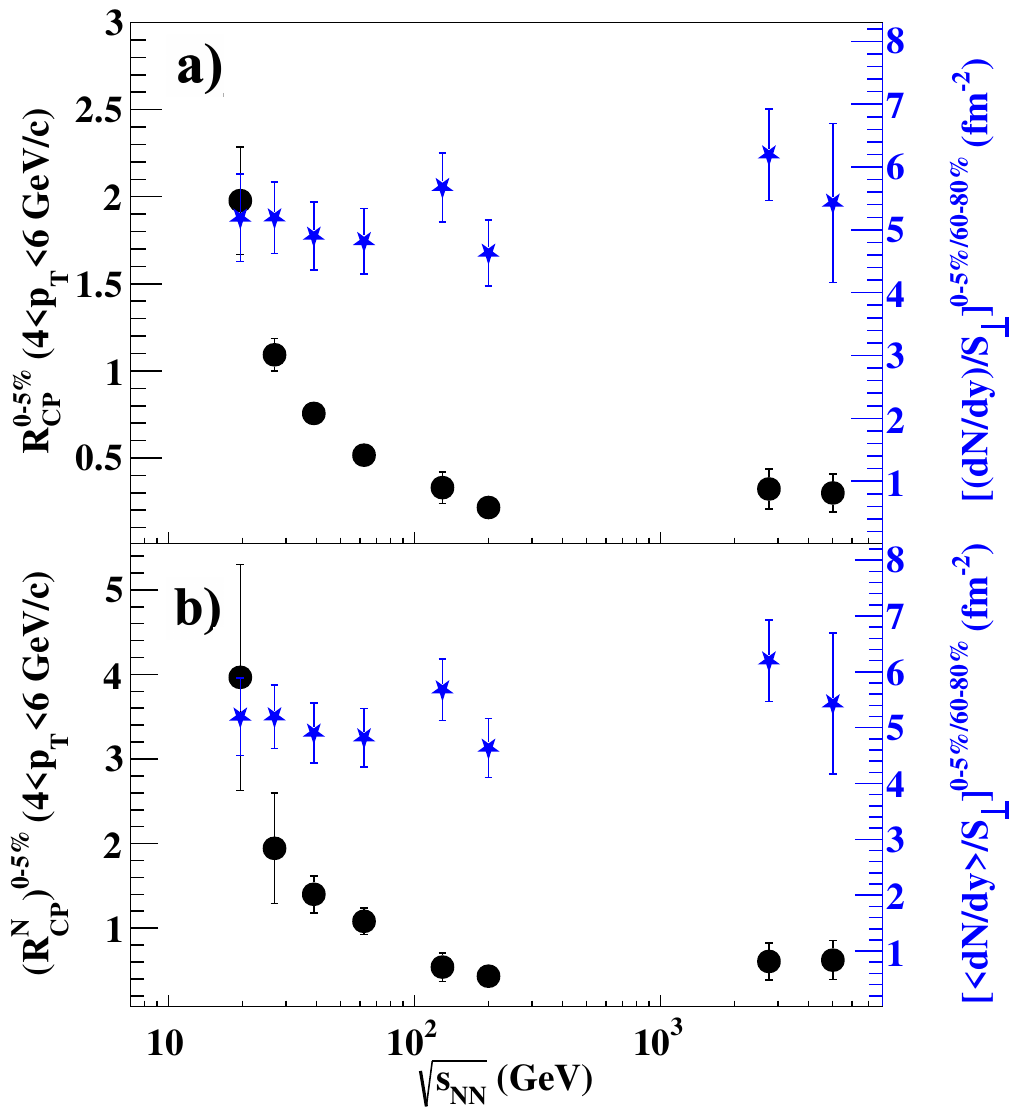}
\caption{a) $R_{CP}$ and b) $R_{CP}^N$, for 4$<p_T<$6 GeV/c, as a function of $\sqrt{s_{NN}}$ for 0-5\% centrality relative to 60-80\%, for charged particles in Au-Au and Pb-Pb collisions. On the right scales, the ratio of particle densities per unit of rapidity and unit of overlapping area for the same centralities is given.}
\label{fig-21}
\end{figure}

 As it is well known, within the Beam Energy Scan (BES) program at RHIC, valuable data were obtained relative to the behaviour of different observables in Au-Au collisions, starting from $\sqrt{s_{NN}}$= 7.7 GeV, up to 39 GeV. Since the $p_T$ spectra for charged particles in pp collisions at these energies were not measured, the STAR Collaboration studied the $p_T$ dependence of $R_{CP}$ [(0-5\%)/(60-80\%)] for different collision energies, for Au-Au collisions \cite{Sangaline12}. In order to include as much as possible the lower energies, where the published data are in a lower $p_T$ range, we had to change the $p_T$ range from 5$<p_T<$8 GeV/c, used in previous sections, to 
4$<p_T<$6 GeV/c, for the study of the charged particle suppression dependence on the collision energy. These results, together with the values obtained in Pb-Pb collisions at $\sqrt{s_{NN}}$=2.76 and 5.02 TeV, for the most central collisions, are presented in Figure \ref{fig-21}a. Following the arguments from the previous section, $R_{CP}^N$ as a function of the collision energy
 is presented in Figure \ref{fig-21}b. 
 In both plots is evidenced a decrease of $R_{CP}$ or $R_{CP}^N$ from 
$\sqrt{s_{NN}}$= 19.6 GeV up to $\sqrt{s_{NN}}$= 200 GeV, while the relative ratios of particle densities per unit of rapidity and unit of overlapping area are constant, within the error bars. Beyond the RHIC energies, $R_{CP}$ and $R_{CP}^N$ remain constant.  
  Since $R_{AA}$ for charged particles at lower RHIC energies are not reported, in order to confirm the above observations, we used the $R_{AA}$ of $\pi^0$ published by the PHENIX collaboration at 
$\sqrt{s_{NN}}$=39, 62.4 and 200 GeV \cite{Adare08, Adare12} and by the ALICE Collaboration \cite{ALICE14, Sekihata18} at LHC energies. 

\begin{figure}[b]
\centering
\includegraphics[scale=0.31]{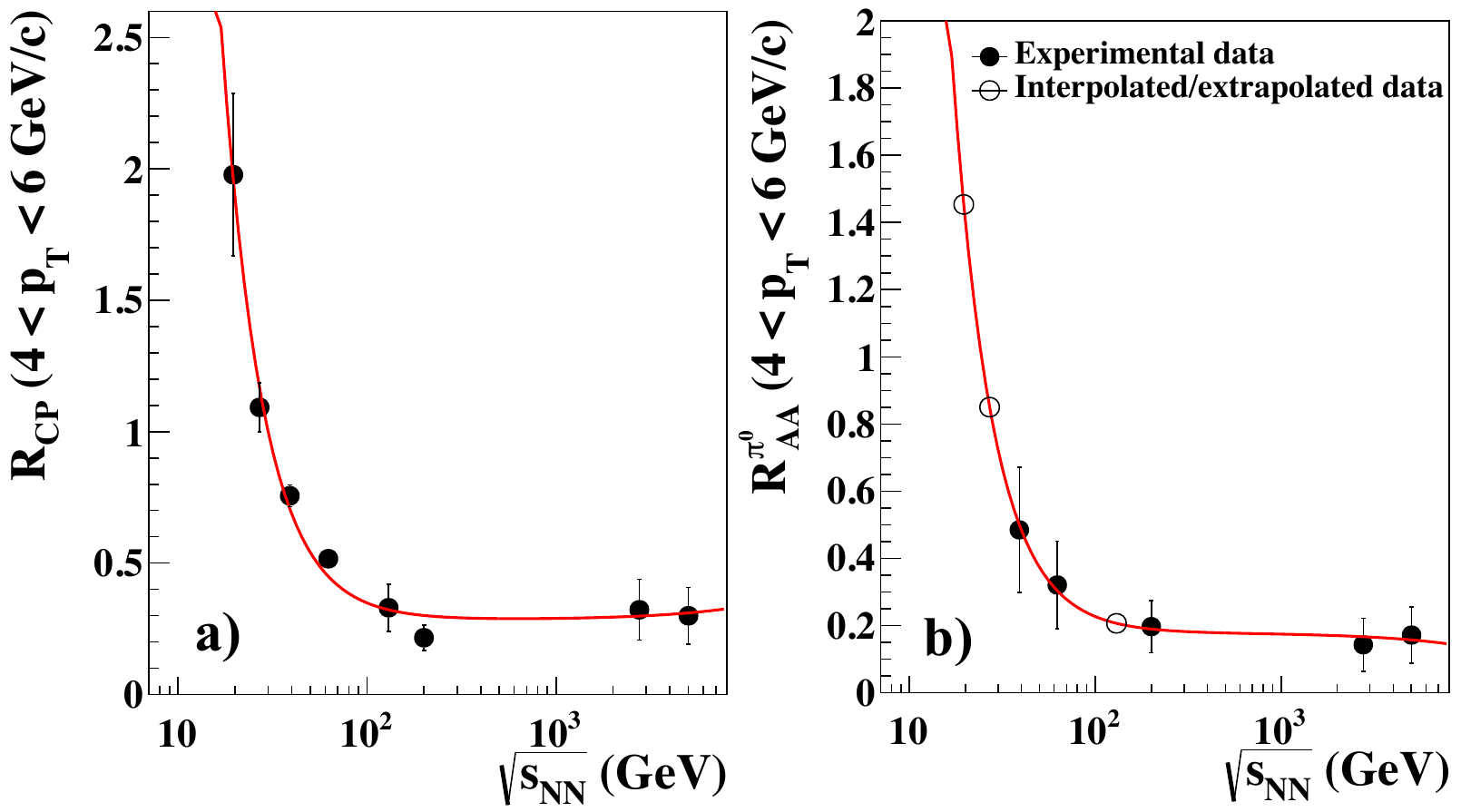}
\caption{a) The same as Figure \ref{fig-21}a. b) $R_{AA}$ for $\pi^{0}$, corresponding to the same range in $p_{T}$ as a), for experimental values (full symbols) and interpolated/extrapolated results (open symbols)  for 0-10\% centrality. In both plots the continuous line is the fit with Eq.15.}
\label{fig-22}
\end{figure}
  
\begin{figure}
\centering
\includegraphics[scale=0.50]{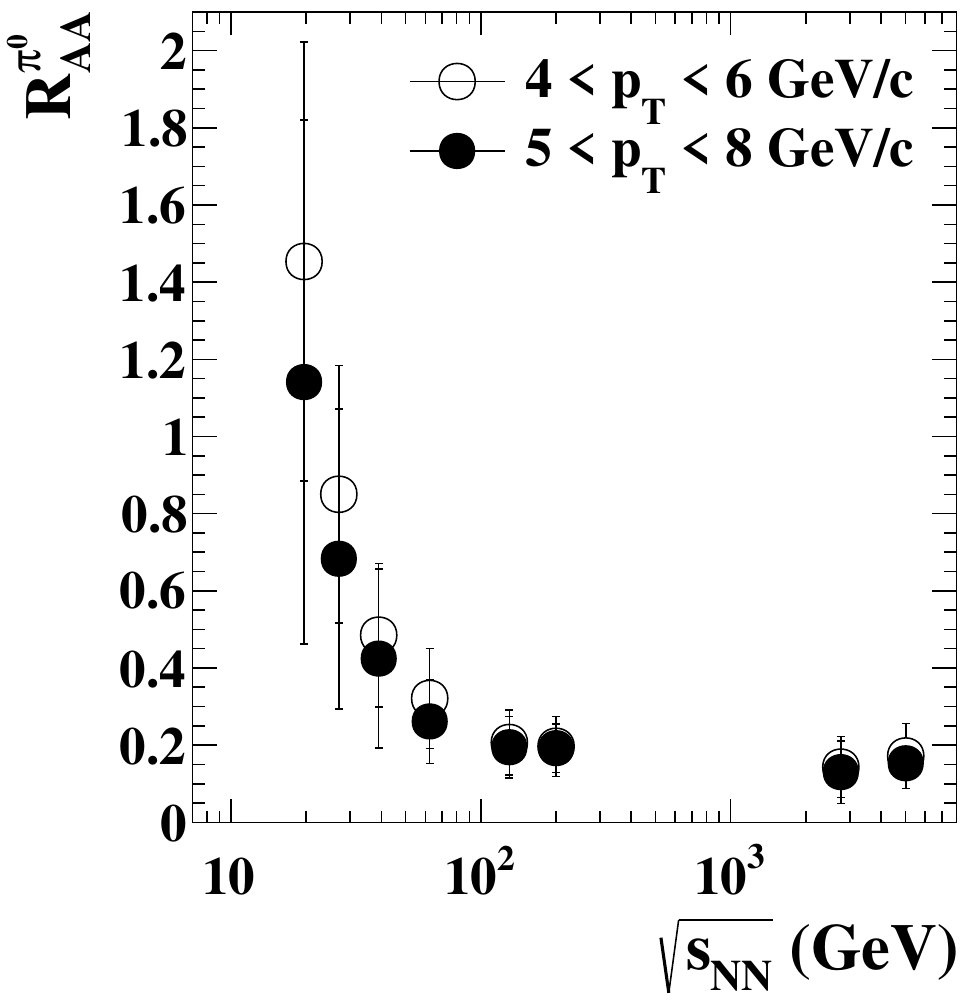}
\caption {$\pi^0$ $R_{AA}$ for the two $p_T$ ranges: 4-6 GeV/c (open symbols) and 5-8 GeV/c (full symbols) for 0-10\% centrality.}
\label{fig-23}
\end{figure}

In order to have an estimate on $R_{AA}^{\pi^{0}}$ corresponding to 0-10\% centrality for the collision energies where it was not published, we applied
the procedure described bellow. 
The  $\sqrt{s_{NN}}$ dependence of $R_{CP}$ (Figure \ref{fig-21}a) was fit with the following empirical expression:
  \begin{equation}
  	R_{CP} \propto a + \frac{b}{s_{NN}} + c\cdot\sqrt{s_{NN}}
  \end{equation}
\noindent  
with a, b, and c as free parameters, the result being presented in Figure \ref{fig-22}a. A similar expression was used in order to fit the measured experimental data of the $R_{AA}^{\pi^{0}}$ - $\sqrt{s_{NN}}$ dependence (Figure \ref{fig-22}b - full symbols), leaving the parameters free. The result was used for estimating $R_{AA}^{\pi^{0}}$ at the missing collision energies, i.e. 19.6, 27 and 130 GeV (Figure \ref{fig-22}b - open symbols). Measured, interpolated and extrapolated $R_{AA}^{\pi^0}$ values as a function of $\sqrt{s_{NN}}$ are presented in Figure \ref{fig-23}, for both $p_T$ ranges used in this paper, namely 4-6 GeV/c (open symbols) and 5-8 GeV/c (full symbols).
 
\begin{figure}
\centering
\includegraphics[scale=0.48]{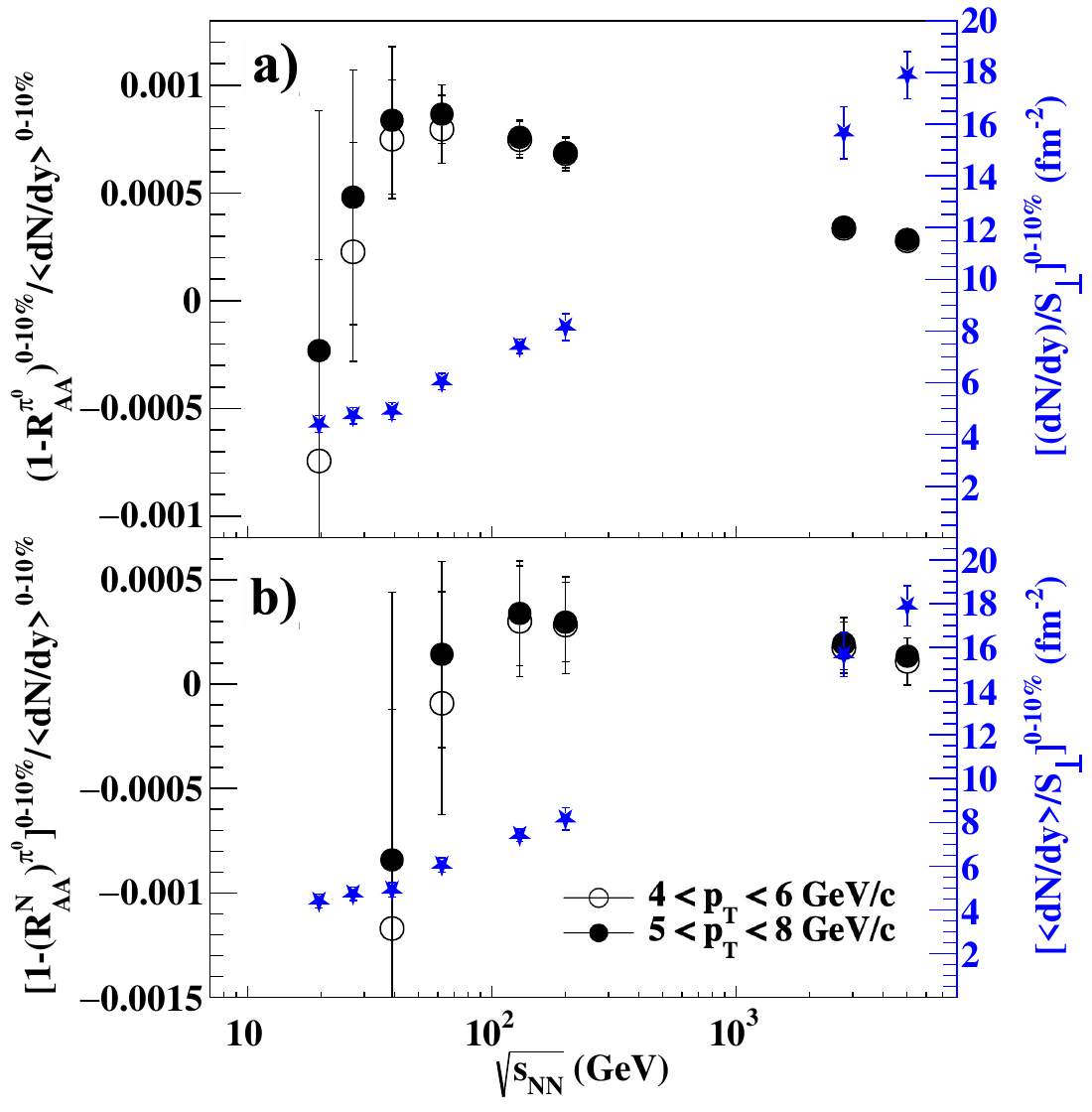}
\caption {a) $(1-R_{AA}^{\pi^0})$/$\langle dN/dy \rangle$ as a function of collision energy; b) $(1-(R_{AA}^N)^{\pi^0})$/$\langle dN/dy \rangle$ as a function of collision energy (bullets)-left scale and ($\langle dN/dy \rangle/S_\perp$) (stars)-right scales for the 0-10\% centrality.}
\label{fig-24}
\end{figure}

 The $R_{AA}^{\pi^0}$ dependence as a function of $\sqrt{s_{NN}}$ is qualitatively similar with the one evidenced for $R_{CP}$ corresponding to charged particles presented in Figure \ref{fig-21}a. The suppression starts around $\sqrt{s_{NN}}$=27 GeV, becomes more significant up to the top RHIC energy and remains constant up to the LHC energies. The ratios relative to $\langle dN/dy \rangle$ as a function of collision energy are presented in Figure \ref{fig-24}, namely: $(1-R_{AA}^{\pi^0})$/$\langle dN/dy \rangle$ (Figure \ref{fig-24}a) and $[1-(R_{AA}^N)^{\pi^0}]$/$\langle dN/dy \rangle$ (Figure \ref{fig-24}b).  
  
These ratios show a maximum around the top RHIC energies (in the region of $\sqrt{s_{NN}}$ = 62.4 - 130 GeV), decreasing towards LHC energies, in qualitative agreement with theoretical predictions \cite{Liao09, Burke14, Shi19}. To what extent such a trend is due to a transition from a magnetic plasma of light monopoles near the critical temperature region \cite{Liao09} to a deconfined matter dominated by quarks and gluons {\cite{Shi19} remains an open question. However, the trends in the experimental data suggest a change in the properties of the deconfined matter from RHIC to LHC energies.

\vspace{1cm}
\section{Conclusions} 
The present paper is mainly based on published experimental data obtained at RHIC and LHC.  The  motivation of this was to study possible scaling or distinctive features between the two energy regimes. Without claiming precise calculations that are extremely laborious, we tried to rely mainly on  experimental considerations.
Based on the experimental results obtained at RHIC for Au-Au, Cu-Cu and at LHC for Pb-Pb and Xe-Xe collisions, a detailed analysis of the charged particle suppression in the region of transverse momentum corresponding to the maximum suppression is presented. 

In order to draw conclusions independent of estimates of the number of binary collisions used in the definitions of $R_{AA}$ and $R_{CP}$, we define the quantities $R_{AA}^N$ and $R_{CP}^N$ in which the ratios of $p_T$ spectra are normalised to charged particle density ($dN_{ch}/d\eta$) before they are then divided by the relevant pp or peripheral $p_T$ spectra, again normalised by charged particle density in pp or peripheral collision.  

While $R_{AA}$ scales as a function of $\langle dN_{ch}/d\eta \rangle$ for the top RHIC and all LHC energies, it scales separately as a function of $\langle N_{part} \rangle$ for RHIC and LHC energies,
for all the corresponding measured colliding systems. 
However, given that $\langle dN_{ch}/d\eta \rangle$ depends on the collision energy and on the overlapping area of the colliding systems, their relative contribution to suppression is rather difficult to unravel. This is the main reason why the considerations on the suppression phenomena as a function of collision geometry and collision energy are mainly based on the $\langle N_{part} \rangle$ dependence. 

The influence of the corona contribution on the experimental $R_{AA}$  is presented. As expected, the main  corona contribution is at low values of $\langle N_{part} \rangle$ where the core suppression relative to the experimental value is larger. 

Based on (1-$R_{AA}$) and $\langle dN/dy \rangle/S_{\perp}\sim T^3$ dependences on $\langle N_{part} \rangle$, one could conclude that a saturation of suppression at LHC energies takes place. At $\langle N_{part} \rangle$=350, corresponding to the most central Au-Au collisions at $\sqrt{s_{NN}}$=200 GeV, if one considers the parton energy loss proportional with the squared path length and with the particle density per unit of rapidity and unit of overlapping area, the proportionality factor $\xi$ is approximately two times lower at LHC than at RHIC. The difference in the hydrodynamic expansion extracted from the $\langle \beta_T \rangle$ scaling as a function of $\sqrt{\langle dN/dy \rangle/S_{\perp}}$ cannot explain this difference. 
Such considerations, applied to the highest charged particle multiplicity measured in pp collisions at 7 TeV could explain why no suppression is evidenced in such events, in the limit of current experimental uncertainties, while there are similarities to Pb-Pb with respect to other observables. $R^N_{AA}$ as a function of $\langle N_{part} \rangle$ shows similar separate scaling for RHIC and LHC energies, with a linear dependence being evidenced. $R^N_{CP}$ shows a very good scaling as a function of $\langle N_{part}\rangle$ for the heavy systems at all collision energies.
The ratio (1-$R_{AA}$)/$\langle dN/dy \rangle$ shows an exponential decrease with 
$(\langle dN/dy \rangle/S_{\perp})^{1/3}$ while  (1-$R_{AA}^N$)/$\langle dN/dy \rangle$ is independent on $(\langle dN/dy \rangle/S_{\perp})^{1/3}$ for $(\langle dN/dy \rangle/S_{\perp})^{1/3}\geq$2.1 particles/fm$^{2/3}$, the value being the same for all the heavy systems at all the collision energies, showing the possible dependence of the jet-medium coupling as a function of temperature.   
For the most central collisions, $R_{CP}$, $R^N_{CP}$ for charged particles and $R_{AA}^{\pi^0}$, $(R_{AA}^N)^{\pi^0}$ for 4$<p_T<$6 GeV/c and 5$<p_T<$8 GeV/c, measured at RHIC in Au-Au collisions and at LHC in Pb-Pb collisions, evidence, as a function of the collision energy, an increase of the suppression from $\sqrt{s_{NN}}$ = 39 GeV up to 200 GeV, followed by a saturation up to the highest energy, $\sqrt{s_{NN}}$ =5.02 TeV for Pb-Pb collisions. $(1-R_{AA}^{\pi^0})$/$\langle dN/dy \rangle$ and $[1-(R_{AA}^N)^{\pi^0}]$/$\langle dN/dy \rangle$ for the 0-10\% centrality evidence a maximum around the largest RHIC energies, in qualitative agreement with models predictions. To what extent this pattern is a signature of a transition in the deconfined matter properties from the top RHIC energy to LHC energies has to be further confirmed by theoretical models. \\
\vspace{-0.7cm}
\section*{ACKNOWLEDGMENTS}
\vspace{-0.2cm}
This work was carried out under the contracts sponsored by the Ministry of Education and Research:  RONIPALICE-04/10.03.2020 (via IFA Coordinating Agency) and PN-19 06 01 03.

\end{document}